\documentclass[preprint,12pt]{elsarticle}

\usepackage{amssymb}
\usepackage{amsmath}
\usepackage{geometry}
\usepackage{mathtools}
\usepackage{mathrsfs}
\usepackage{bm}
\usepackage{graphicx}
\usepackage{indentfirst}
\usepackage{xcolor}
\usepackage{fancyhdr}
\usepackage{float}
\usepackage[normalem]{ulem}

\journal{Chemical Engineering Journal}

\begin{document}

\begin{frontmatter}

\title{Rethinking Balance Sheets: A Poisson-Nernst-Planck Based Approach for Modeling Concentration and Flux Profiles Inside an Electrochemical Cell} 

\author[a]{Grace Origer}
\author[a]{Ritu R. Raj}
\author[e]{Nathan Jarvey}
\author[a]{P. N. Romero Zavala}
\author[a,d]{Wilson A. Smith}
\author[a,b,c]{Ankur Gupta \corref{cor1}}

\affiliation[a]{organization={Department of Chemical and Biological Engineering, University of Colorado Boulder},
            city={Boulder},
            postcode={80303}, 
            state={CO},
            country={U.S.A}}
\affiliation[b]{organization={Materials Science and Engineering Program, University of Colorado Boulder},
            city={Boulder},
            postcode={80303}, 
            state={CO},
            country={U.S.A}}
\affiliation[c]{organization={Department of Applied Mathematics, University of Colorado Boulder},
            city={Boulder},
            postcode={80303}, 
            state={CO},
            country={U.S.A}}

\affiliation[d]{organization={Renewable and Sustainable Energy Institute, University of Colorado Boulder},
            city={Boulder},
            postcode={80303}, 
            state={CO},
            country={U.S.A}}

\affiliation[e]{organization={Chemistry and Nanoscience Center, National Laboratory of the Rockies},
            city={Golden},
            postcode={80401}, 
            state={CO},
            country={U.S.A}}
            
\cortext[cor1]{ankur.gupta@colorado.edu}
\begin{abstract}
Electrochemical cells serve as a building block for producing and storing electrical energy from chemical reactions. The analysis of ion transport in these systems forms the foundation for understanding more complex electrochemical systems that are becoming increasingly present in the broader societal energy infrastructure. From a pedagogical perspective, the ``balance sheets" introduced in Chapter 4 of Electrochemical Methods: Fundamentals and Applications by Alan J. Bard, Larry R. Faulkner  and Henry S. White (hereafter referred to as BFW) provides a first-pass approach to analyze ion transport in electrochemical cells. However, the balance sheet approach lacks first-principles justifications from the underlying equations that describe the transport processes in electrochemical cells. In this work, we compare a first-principles approach via the Poisson-Nernst-Planck equations to describe ion transport in electrochemical cells to that of the balance sheet approach. By re-working the  examples presented in BFW, we illustrate that the balance sheet approach is only valid in limited scenarios. Furthermore, we show that the PNP equations provide a more physical route to analyze ion transport in electrochemical systems. We hope the approach outlined here will be adopted by  electrochemical engineering researchers and instructors.
\end{abstract}




\end{frontmatter}


\section{Introduction}
Electrochemical cells serve as fundamental building blocks for a range of modern electrochemical systems. They are used in energy storage devices \cite{maddukuri_challenge_2020,bgoodenough_electrochemical_2014} such as batteries and fuel cells \cite{armand_building_2008,lucia_overview_2014} and in electrolytic applications, such as water electrolysis \cite{shih_water_2022,grigoriev_current_2020}, electroplating \cite{gugua_electroplating_2024}, and metal refining \cite{jin_sustainable_2020,su_electrochemical_2020}. In an electrochemical cell, ions in an electrolyte solution participate in reduction-oxidation (redox) reactions at electrode surfaces to produce a current under an applied electric field.
One aspect of electrochemical cell performance, of interest to this work, are the mass transfer mechanisms responsible for ion transport through the bulk electrolyte. In order for the surface (Faradaic) reactions to take place, the ions need to transport to their appropriate electrodes. When neglecting convective transport, electrolyte mass transfer occurs via two mechanisms: diffusion and electromigration \cite{tobias_fiftieth_1952, bard2022electrochemical}. Diffusion refers to mass transfer driven by ion concentration gradients created by spatial variations in ion concentrations. Electromigration refers to the transport of ions by a net force due to an applied electric field. The total current generated by the electrochemical surface reactions is carried from one electrode to the other via these two mass (species) transport mechanisms \cite{dickinson_comsol_2014}.

In the case of a mass transfer-limited reaction, where the Faradaic reactions can be treated as instantaneous, the rate of electrolysis in the cell is limited by the rate at which ions can be transported to (or away from) the electrode by diffusion and electromigration \cite{tobias_fiftieth_1952}. In this scenario, the current cannot exceed a maximum limiting current. The limiting current is set by the maximum rate at which ion depletion can be counteracted by the transport of new ions to the surface \cite{selman_mass-transfer_1978}.  In this case, current is limited by the ion mass transfer rate. In contrast, for finite reaction rates corresponding to currents below the limiting value, the reaction rate balances the species transport to the electrode to set the current \cite{charlot_electrochemical_1962}. Developing a proper understanding of mass transfer in electrochemical cells, particularly for mass transfer limited systems, can be vital in determining the performance of an electrochemical cell.

Electrochemical Methods: Fundamentals and Applications by Alan J. Bard, Larry R. Faulkner and Henry S. White (hereafter referred to as BFW) is a seminal textbook for understanding basic electrochemical phenomena, including ion transport \cite{bard2022electrochemical}. In Chapter 4, to understand the contribution of various transport phenomena to the cell current, BFW uses a ``balance sheet" analysis (hereafter referred to as BSA) approach to visually decompose an arbitrary reaction current into its diffusive and electromigrative components \cite{kolthoff_polarography_1952,koryta_electrochemistry_1970,charlot_electrochemical_1962,bard2022electrochemical}. In the BSA analysis, species concentrations are assumed to be uniform throughout the electrochemical cell, except in a small region near each electrode. This concentration assumption forces electromigration to be the dominant transport mechanism in the bulk of the cell and only allows diffusion to occur in a small region near the electrodes. The electromigrative contribution of each species to the total current is determined in this bulk region via the use of equivalent conductance and transference numbers \cite{koryta_electrochemistry_1970, bard2022electrochemical}. Following this, the diffusive contribution to current near the electrodes can be determined by enforcing species conservation given the imposed surface reaction rates \cite{bard2022electrochemical}. To see an example of the BSA formulation, see Section 2.1 of this article. BFW uses this BSA approach to analyze the contribution of specific transport mechanisms to the overall current in three examples: i) a copper redox cell without and ii) with a supporting electrolyte and iii) in a hydrogen evolution cell.
While BSA provide a first-pass approach for analyzing mass transfer in an electrochemical cell, a number of \textit{ad-hoc} assumptions are invoked. First, the approach assumes that there are two regions: a bulk region and a boundary layer region. The bulk region is assumed to have no concentration gradients, and hence  the diffusive fluxes in the bulk are zero. This assumption is made without reference to mixing conditions, specific timescale, nor justification based on analysis of governing transport equations, such as the Poisson-Nernst-Planck (PNP) equations \cite{bard2022electrochemical}. A lack of physical justification makes it difficult to understand under what scenarios, even for the examples posed in the text, the analysis would be valid. Second, the diffusive fluxes, by definition of the approach, are discontinuous between the bulk and electrode region. This means that each species is not conserved across the cell, breaking mass conservation. Third, despite the presence of concentration gradients near the electrode, the electromigrative flux is assumed to remain at the bulk value near the electrodes. This is inconsistent since the bulk electromigrative flux values were determined assuming zero concentration gradients. Fourth, there are no considerations given to the amount of current flowing through compared to the limiting current. Lastly, there is no prediction of the structure of the species concentration profiles, making it difficult to determine if the above assumptions are valid \textit{post-hoc}. 

In this article, we assess the validity of the BSA approach to study mass transfer in electrochemical cells by re-analyzing the examples outlined in BFW through the framework of the PNP equations. The PNP equations are a set of continuum governing ionic species transport equations that contain a species balance for each ion coupled with Poisson's equation for the electric potential \cite{sharifi_golru_influence_2024,bui_dynamic_2021,bazant2004diffuse, kilic_steric_2007,schmuck_homogenization_2015,jarvey_asymmetric_2023,jarvey_ion_2022, henrique2022charging}. By conducting such an analysis, we develop a first-principles approach to calculate the concentration and flux profiles for each species in a given electrochemical cell. We note that by solving the PNP equations by itself is not new (see Newman and Balsara for a thorough in-depth analysis~\cite{newman2021electrochemical}), but to the best of our knowledge, the scenario for the BSA presented in BFW have not been analyzed systematically. By solving the PNP equations directly, we will systematically compare our calculated contributions to the overall current to those determined via the BSA approach. Specifically, we shed light on assumptions made through the BSA approach and provide a first-principles framework for analyzing mass transport in these instructive electrochemical cell models. We note that PNP equations themselves may need to be modified to account for a variety of effects such as finite ion-size~\cite{kilic_steric_2007, borukhov2000adsorption, gupta2018electrical}, dielectric decrement~\cite{nakayama2015differential, gupta2018electrical} and ion-ion correlations~\cite{bazant2011double, gupta2020ionic, de2020interfacial}, among others. Nonetheless, the concepts outlined here remain valid.

The goals of this article are two-fold. 1) We evaluate the correctness of the flux predictions made by the BSA approach via comparison to the prediction of the PNP equations. By doing so, we provide clarity for the physical assumptions used in the analysis and determine if the predictions made by BSA are correct under those assumptions. 2) We demonstrate how relatively accessible mathematical and computational tools can be used to solve governing equations for ion transport. From a pedagogical perspective, we hope this work will provide instructors with new tools to implement in the classroom that enables better clarity with regard to teaching mass transfer in electrochemical cells. From a research perspective, we believe that our approach helps researchers explore new mechanisms for modeling that will aid them in building first-principles backed explanations of their observations.

We begin, in Section 2, by re-analyzing the copper redox reaction without a supporting electrolyte. The section starts by summarizing the BSA approach to predict the diffusive and electromigrative contributions to the current. Next, we analyze the system through the lens of the PNP equations. The resulting steady-state simulations of concentration and flux profiles are compared to BSA predictions. Following the steady-state analysis, we solve the full transient problem to probe the validity of the BSA method as the system approaches steady state. Lastly, for completeness, we conduct an analysis assuming bulk mixing of the electrochemical cell and compare with the BSA predictions. In Section 3, we repeat the analysis for the copper redox cell that includes a supporting electrolyte. Lastly, in Section 4, we conduct a similarity variable analysis for the hydrogen evolution cell.

\section{Analysis of a Copper Redox Cell}
\begin{figure}[!ht]
\centering
\includegraphics[width=5.26in]{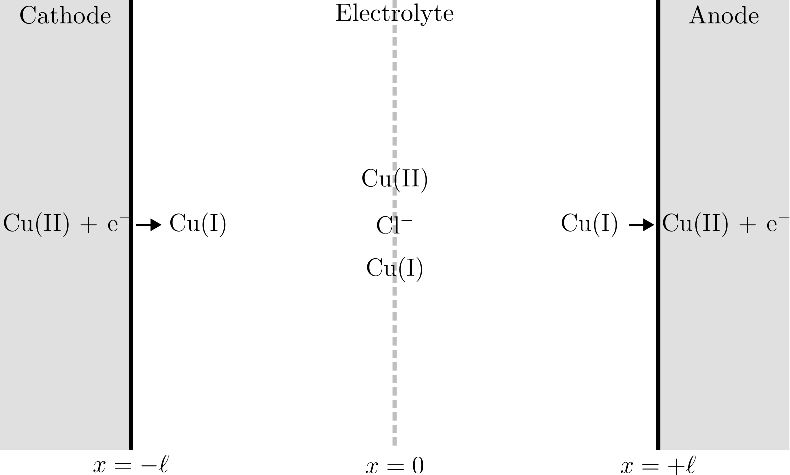}
\caption{\textbf{Schematic representation of copper redox cell presented, as given in Section 4.3 of Bard, Faulkner and White \cite{bard2022electrochemical}.} An electrolyte solution composed of $10^{-3}$ M Cu(NH$_3)_4^{2+}$, $10^{-3}$ M Cu(NH$_3)_2^{+}$, and $3 \times 10^{-3}$ M Cl$^-$ in $0.1$ M NH$_3$ is placed in an electrochemical cell with a cathode at $x = -\ell$ and an anode at $x = \ell$. A current is produced via the reduction of Cu(II) at  the cathode and the oxidation of Cu(I) at the electrode.}
\label{fig1}
\end{figure}

We first analyze the copper redox cell discussed in Figure 4.3.3 in BFW \cite{bard2022electrochemical}. In this example, a one-dimensional electrochemical cell is loaded with $10^{-3}$ M Cu(NH$_3)_4^{2+}$, $10^{-3}$ M Cu(NH$_3)_2^{+}$, and $3\times10^{-3}$ M Cl$^-$ in $0.1$ M NH$_3$, see Fig. \ref{fig1}. At the left electrode, located at $x = -\ell$, Cu(II) is reduced to Cu(I) by reacting with electrons. At the right electrode, located at $x = \ell$, Cu(I) is oxidized to Cu(II), thereby producing electrons in the process. $k$ denotes a turnover rate which is equivalent to the number of electrons reacted per unit time, and is assumed to be constant. As the reaction proceeds, the concentration of Cu(I) and Cu(II) is either locally enriched or depleted, producing a concentration gradient driving a diffusive flux. In addition, an electric field is produced between the cathode and anode which drives an electromigrative flux of the ionic species between the electrodes. 
The diffusive and electromigrative fluxes work in tandem to ensure that species and current conservation are maintained throughout the electrochemical cell. A steady state is achieved when fluxes of Cu(II) and Cu(I) become constant. We provide a summary of the mathematical symbols used in Table~\ref{tab: symbols}.

Similar to the example in BFW, we assume that the limiting equivalent conductance ($\lambda_{g}$), i.e. the equivalent conductance at infinite dilution, of all $g^{\textrm{th}}$ species are equal \cite{kolthoff_polarography_1952,bard2022electrochemical}. As such: $\lambda_{\textrm{Cu(II)}} = \lambda_{\textrm{Cu(I)}} = \lambda_{\textrm{Cl$^-$}} = \lambda$. By definition
\begin{equation}
    \lambda_{g} = \frac{F^2}{RT} |z_{g}|D_{g},
    \label{eqn:lim_cond}
\end{equation}
where $F$ is Faraday's constant, $R$ is the universal gas constant , $T$ is the absolute temperature of the electrolyte, $z_{g}$ is the valence of the $g^{\textrm{th}}$ species, and $D_{g}$ is the diffusivity of the $g^{\textrm{th}}$ species. Assuming $\lambda_{g}$ to be the same for all ionic species is equivalent to writing
\begin{equation}
\begin{matrix}
    D_{\textrm{Cu(II)}} \\ 
    D_{\textrm{Cu(I)}} \\
    D_{\textrm{Cl}^{-}} 
\end{matrix} = 
\begin{matrix}
    \frac{D}{2}, \\ D, \\ D.
\end{matrix}
\label{eqn:copper_redox_diffusivity}
\end{equation}

\subsection{Balance Sheet Approach} 

\begin{table}[tb!]
    \centering
    \caption{Description of mathematical variables and corresponding units}
    \label{tab: symbols}
    
    \begin{tabular}{|l|c|l|} 
        \hline 
        \textbf{variable} & \textbf{units} & \textbf{description} \\
        \hline 

        $R$ & J/mol/K & universal gas constant \\
        \hline

        $T$ & $K$ & temperature \\
        \hline

        $F$ & coulomb/mol & Faraday's constant \\
        \hline
        
        $k$ & \#/s & turnover rate \\
        \hline

        $\phi$ & V & electric potential \\
        \hline

        $x$ & m & coordinate \\
        \hline
        
        $A_\textrm{electrode}$ & m$^2$ & area of the electrode \\
        \hline
  
        $N_A$ & \#/mol & Avogadro's constant\\
        \hline
        
        $R_\textrm{V,i}$ & mol/m$^3$/s & volumetric reaction rate \\
        \hline 

        $i$ & \#/s & total number of electrons per unit time transported \\
        \hline 

        $i_d$ & \#/s & number of electrons per unit time transported by diffusion \\
        \hline

        $i_m$ & \#/s & number of electrons per unit time transported by electromigration \\
        \hline
        
        $n_g$ & \#/s & molecular transport rate of the $g^{\textrm{th}}$ ion \\
        \hline

        $n_{d,g}$ & \#/s & molecular transport rate of the $g^{\textrm{th}}$ ion due to diffusion \\
        \hline

        $n_{m,g}$ & \#/m$^2$/s & molecular transport rate of the $g^{\textrm{th}}$ ion due to electromigration\\
        \hline
        
        $c_g$ & \#/m$^3$ & atomic concentration \\
        \hline
              
        $D_g$ & m$^2$/s & diffusivity of $g^\textrm{th}$ species \\
        \hline
        
        $t_g$ & dimensionless & transference number of $g^\textrm{th}$ species \\
        \hline
        
        $C_g$ & mol/m$^3$ & molar concentration of $g^\textrm{th}$ species, $C_g = c_g/N_A$\\
        \hline

        $N_g$ & mol/m$^2$/s & molar flux of g$^\textrm{th}$ species \\
        \hline 

         $z_g$ & dimensionless & valence of g$^\textrm{th}$ species \\
         
        \hline 

        $\nu$ & coulomb/m$^2$/s & charge flux at the electrode\\
        \hline

        $\varepsilon$ & Farad/m & dielectric permittivity of the medium \\
        \hline

        $e$ & coulomb & elementary charge of an electron \\
        \hline

        $j$ & coulomb/m$^2$/s & total charge flux \\
        \hline
    \end{tabular}
\end{table}

In this section, we will outline the BSA analysis which has two main goals. The first is to demonstrate the claim that during electrolysis, electromigration is primarily responsible for the transport of current in the bulk solution, while diffusion occurs only in the vicinity of the electrode. The second is to evaluate which species are primarily responsible for carrying the current in the various spatial regions of the cell.

In the copper redox cell depicted in Fig. \ref{fig1}, BFW assume a turnover rate of $k=6$ electrons per unit time at each electrode. This implies that at the cathode, 6 Cu(II) atoms per unit time are reduced to 6 Cu(I) atoms. At the anode, 6 Cu(I) atoms per unit time are oxidized to 6 Cu(II) atoms. Cl$^-$ is assumed to be inert and is needed to balance the positive copper ions. We define transport rate $n_{g}$ (unit - \# atoms/second) at which the $g^{\textrm{th}}$ charged species migrate by both diffusion and electromigration. By definition, the current in the system, in units of number of electrons per unit time
\begin{equation}
    i = \sum_{g}z_{g}n_{g}.
    \label{equn:atomic_current_def}
\end{equation}
We show later that this current must remain constant if electroneutrality is imposed. For instance, at the left electrode, $i = (z_{\textrm{Cu(II)}}n_{\textrm{Cu(II)}}+z_{\textrm{Cu(I)}}n_{\textrm{Cu(I)}}+z_{\textrm{Cl}^{-}}n_{\textrm{Cl}^{-}}) = (2\times(-6) + 1\times6 + (-1)\times0) = -6$. There is a current of $-6$ electrons per unit time flowing left to right in the cell or $6$ electrons per unit time flowing right to left in the cell. This atomic rate of each individual species can be expressed in terms of diffusive and electromigrative components as

\begin{equation}
    n_{g} =-\underbrace{A_{\textrm{electrode}}D_{g}\frac{\partial c_{g}}{\partial x}}_{n_{d,g}} - \underbrace{ A_{\textrm{electrode}}\frac{F}{RT}z_{g}D_{g}c_{g}\frac{\partial \phi}{\partial x}}_{n_{m,g}},
    \label{equn:atomic_rate}
\end{equation}
where $A_{\textrm{electrode}}$ is the area of the electrode, $c_{g}$ is the atomic concentration of the $g^{\textrm{th}}$ species, and $\phi$ is the electric potential. We can decompose the total current $i$ into the diffusive $i_d$ and electromigrative $i_m$ components as

\begin{equation}
    i = i_d + i_m.
    \label{equn:current_balance}
\end{equation}
Using eqn. \ref{equn:atomic_current_def} and eqn. \ref{equn:atomic_rate}, we write
\begin{equation}
    i = -\underbrace{\sum_{g}A_{\textrm{electrode}}z_{g}D_{g}\frac{\partial c_{g}}{\partial x}}_{i_d} - \underbrace{\sum_{g}A_{\textrm{electrode}}\frac{F}{RT}z^2_{g}D_{g}c_{g}\frac{\partial \phi}{\partial x}}_{i_m}.
    \label{eqn:bulk}
\end{equation}
In the bulk region, since $\frac{\partial c_{g}}{\partial x} = 0$, Eq. \ref{eqn:bulk} becomes

\begin{equation}
    i = -\sum_{g}A_{\textrm{electrode}}\frac{F}{RT}z^2_{g}D_{g}c_{g}\frac{\partial \phi}{\partial x}.
    \label{Eq: bulk}
\end{equation}

Using  Eq.~\ref{Eq: bulk}, the electromigrative contribution to the current in bulk from each species $i_{m,g}$ can be determined as

\begin{equation}
    i_{m,g} = t_{g}i,
    \label{eqn:em_current}
\end{equation}
where $t_{g}$ is defined as the transference number and is written as
\begin{equation}
    t_{g} = \frac{z^2_{g}D_{g}c_{g}}{\sum_{g}z^2_{g}D_{g}c_{g}}.
    \label{eqn:transference_number}
\end{equation} 
We emphasize, that while BFW define $t_{g}$ as the fraction of total current that the g$^{\textrm{th}}$ ion carries, the definition provided above is only valid assuming no concentration gradients in the bulk. If there is diffusion in the bulk electrolyte, for instance due to concentration gradients, the above definition for transference number cannot be used, as also acknowledged by BFW. Using the transference number as defined in eqn. \ref{eqn:transference_number}, we find that the electromigrative contribution to the current from each species as

\begin{equation}
\begin{matrix}
    i_{m,\textrm{Cu(II)}} \\ 
    i_{m,\textrm{Cu(I)}} \\
    i_{m,\textrm{Cl}^{-}} 
\end{matrix} = 
\begin{matrix}
    -2, \\ -1, \\ -3.
\end{matrix}.
\label{eqn:copper_redox_current}
\end{equation}
We can then write the atomic rate of each species in the bulk, which is assumed to be due only to electromigration, using eqn \ref{equn:atomic_current_def} as

\begin{equation}
\begin{matrix}
    n_{\textrm{Cu(II)}} \\ 
    n_{\textrm{Cu(I)}} \\
    n_{\textrm{Cl}^{-}} 
\end{matrix} = 
\begin{matrix}
    -1, \\ -1, \\ 3.
\end{matrix}.
\end{equation}

One can immediately notice that the species atomic rates in the bulk do not balance those at the electrodes. While this is not necessarily an issue for a transient system or a well-mixed system, as we will show later, at steady-state without mixing, the total atomic rate of each species throughout the system must spatially be constant. 

The next step in the BSA approach is to balance atomic rates near the electrodes with (to be determined) diffusive rate. So far, the electromigrative rate has been the sole carrier of species transport in the bulk and does not equal total reaction rate at the electrodes, determined by the surface reactions. Therefore, diffusion is used to balance the species rates to their respective surface reaction rates. In these regions, it is assumed that the electromigrative rate remains constant at the bulk values. We show later why this might not be the case. 
Nonetheless, at the anode, $n_{g}$ is based on the Faradaic reaction

\begin{equation}
\begin{matrix}
    n_{\textrm{Cu(II)}} \\ 
    n_{\textrm{Cu(I)}} \\
    n_{\textrm{Cl}^{-}} 
\end{matrix} = 
\begin{matrix}
    -6 \\ 6 \\ 0
\end{matrix}.
\end{equation}

We then write $n_{d,g} = n_{g} - n_{m,g}$, which allows us to solve for the diffusive rate at the anode as

\begin{equation}
\begin{matrix}
    n_{\textrm{d,Cu(II)}} \\ 
    n_{\textrm{d,Cu(I)}} \\
    n_{\textrm{d,Cl}^{-}} 
\end{matrix} = 
\begin{matrix}
    -5 \\ 7 \\ -3
\end{matrix}.
\end{equation}

\begin{figure}[!ht]
\centering
\includegraphics[width=5.1in]{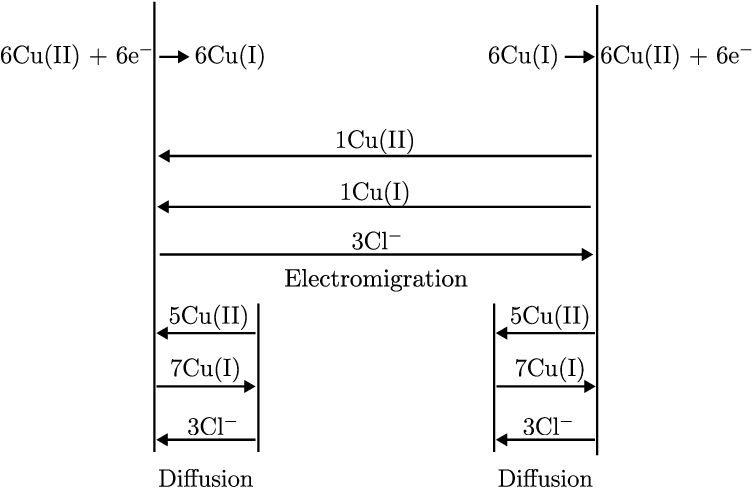}
\caption{\textbf{Recreation of Figure 4.3.3 from Bard, Faulkner and White \cite{bard2022electrochemical}.} Depiction of electromigrative and diffusive rate contributions as determined by the balance sheet procedure.}
\label{fig2}
\end{figure}

A similar procedure can be performed at the cathode. A graphic summary of the rates determined via the above procedure can be seen in Fig. \ref{fig2}.

The BSA provides a relatively straightforward approach that only requires basic arithmetic to predict the contributions of diffusion and electromigration to the current. However, the process through which the BSA approach determines these rates is inconsistent. Ultimately, there are two key assumptions that remain unaddressed: 1) How does one know that the concentration profiles are gradient-free far away from the electrode without assumptions of a well-mixed and/or unsteady conditions? Furthermore, if the system is assumed to be at steady state, why dont the species transport rates in the bulk match the surface reaction rate values at the electrodes? It might be argued that the diffusive rate from both electrodes is what provides the species balance in the bulk. However, if the bulk is gradient-free, there indeed cannot be a diffusive rate in the bulk. 2) When the electromigrative rate is determined in the bulk, how can it be assumed to remain at the bulk value in a region now with a concentration gradient? Throughout the next section we will re-analyze this problem from the perspective of the PNP equations. We discuss the validity of some assumptions and clarify areas of confusion.

\subsection{Analysis via Poisson Nernst Planck (PNP) Equations}
\subsubsection{Problem Set Up}

We compare the results from the BSA approach directly to the solutions of the PNP equations. 
The PNP equations describe the conservation for the $g^{\textrm{th}}$ species as
\begin{equation}
    \underbrace{\frac{\partial C_g}{\partial t}}_{\textrm{accumulation}} = \underbrace{-\bm{\nabla}\cdot \bm{N}_g}_\textrm{in-out} + \underbrace{R_{\textrm{V}_g}}_\textrm{reaction},
\label{eqn:species_transport}
\end{equation}
where $C_g$, $R_{\textrm{V}g}$, and $\bm{N}_g$ are the molar concentration, the volumetric reaction rate, and the molar flux of the $g^{\textrm{th}}$ species respectively and $\bm{\nabla} = \begin{bmatrix}
    \frac{\partial}{\partial x},\frac{\partial}{\partial y},\frac{\partial}{\partial x}
\end{bmatrix}$ is the gradient operator. The molar concentration is related to the atomic concentration via $C_{g} = c_{g}/N_{\textrm{A}}$ where $N_{\textrm{A}}$ is the Avogadro's number. Similarly, the molar flux $\bm{N}_g$ is related to the vector atomic rate $\bm{n}_g$ via $\bm{N}_g =  \bm{n}_g /\left(A_{\textrm{electrode}}\times N_{\textrm{A}}\right)$. The species balances are solved simultaneously with the Poisson equation 
\begin{equation}
    -\varepsilon\nabla^2\phi = F\sum_gz_gC_g,
\label{eqn:vector_poisson_eqn}
\end{equation}
where $\varepsilon$ is the dielectric permittivity of the medium. We consider a one-dimensional analysis of the electrochemical cell and neglect volumetric reactions. We write eqns. \ref{eqn:species_transport} and \ref{eqn:vector_poisson_eqn} as
\begin{equation}
    \frac{\partial C_g}{\partial t} = -\frac{\partial}{\partial x}N_{\textrm{g,}x} = D_g\frac{\partial^2C_g}{\partial x^2} + z_gD_g\frac{F}{RT}\frac{\partial}{\partial x}\left(C_g\frac{\partial \phi}{\partial x}\right),
\label{eqn:species_transport_1D}
\end{equation}
and
\begin{equation}
    -\varepsilon\frac{\partial^2 \phi}{\partial x^2} = F\sum_gz_gC_g.
\label{eqn:poisson_eqn_1D}
\end{equation}
 
For our analysis we do not consider the presence of electrical double layers, allowing us to reduce eqn. \ref{eqn:poisson_eqn_1D} to 
\begin{equation}
    \sum_gz_gC_g = 0,
\label{eqn:electroneutrality}
\end{equation}
 which is known as the so-called electroneutrality condition. We note that some reports incorrectly assume that electroneutrality requires the electric field to be constant. However, this is inaccurate since upon non-dimensionalization, the pre-factor on left hand-side of Eq.~\ref{eqn:poisson_eqn_1D} vanishes far away from the electrode~\cite{deen_analysis_2012}. Therefore, even under the electroneutrality assumption, electric field can vary.   
 
 We solve eqns. \ref{eqn:species_transport_1D} and \ref{eqn:electroneutrality} in the geometry depicted in Fig. \ref{fig1} to determine $C_{g}$ and $N_{g,x}$ for each species. For simplicity, as we only consider one-dimensional systems, for the remainder of this work we write $N_{g,x}$ as $N_{g}$. Before solving, we define the boundary conditions and initial conditions. We recall that $k$ is the number of electrons passed at the electrode per unit time. This allows us to define charge flux at the electrode as $\nu = \frac{ke}{A_{\textrm{electrode}}}$, where $A_{\textrm{electrode}}$ is the area of the electrode and $e$ is the elementary charge of the electron. We then define our boundary conditions as $N_{\textrm{Cu(II)}}(x = \pm \ell) = -\nu/F$, $N_{\textrm{Cu(I)}}(x = \pm \ell) = \nu/F$, and $N_{\textrm{Cl$^{-}$}}(x = \pm \ell) = 0$ as well as the initial conditions $C_{\textrm{Cu(II)}}(x,t = 0) = C_0$, $C_{\textrm{Cu(I)}}(x,t = 0) = C_0$, and $C_{\textrm{Cl$^{-}$}}(x,t = 0) = 3C_0$.

To avoid having to solve for $\phi$ via the constraint of electroneutrality, eqns. \ref{eqn:species_transport_1D} and \ref{eqn:electroneutrality} can be manipulated to eliminate $\frac{d\phi}{dx}$ by multiplying eqn. \ref{eqn:species_transport_1D} by $z_g$ and summing over i to write

\begin{equation}
    \frac{\partial}{\partial t} \sum_g z_gC_g = -\frac{\partial}{\partial x}\sum_g z_gN_g = -\frac{1}{F}\frac{\partial}{\partial x}{j},
\label{eqn:current_conversation}
\end{equation}
where $j$ is the total current density in the system. 
Using electroneutrality, we write
\begin{equation}
    -\frac{1}{F}\frac{\partial}{\partial x}{j} = 0 \implies j = F\sum_g z_gN_g = \sum_g z_gD_g\frac{\partial C_g}{\partial x} + \sum_gz_g^2D_g\frac{F}{RT}C_g\frac{\partial \phi}{\partial x} =\textrm{constant}.
\label{eqn:constant_current}
\end{equation}
This implies that the current in the system is independent of time and space. In this case, we evaluate the current at $x = -\ell$ and find that $j$ = -$\nu$. However, the specific contributions of diffusion and electromigration, in addition to the temporal evolution of those contributions, remain to be solved. We then rearrange eqn. \ref{eqn:constant_current} to solve for $\frac{\partial \phi}{\partial x}$ as
\begin{equation}
    \frac{\partial \phi}{\partial x} = -\frac{-\nu + F\sum\limits_g z_gD_g\frac{\partial C_g}{\partial x}}{\frac{F^2}{RT}\sum\limits_{g}z_g^2D_gC_g}.
\label{eqn:solve_for_dphi_dx}
\end{equation}
Eqns. \ref{eqn:species_transport_1D} can now be solved with eqn. \ref{eqn:solve_for_dphi_dx}, along with the boundary and initial conditions, to determine $C_g(x,t)$ for each species. Before solving, we non-dimensionalize the equations as
\begin{equation}
    \frac{\partial \tilde{C}_g}{\partial \tau} = \mathscr{D}_g\frac{\partial^2\tilde{C}_g}{\partial \tilde{x}^2} + z_g\mathscr{D}_g\frac{\partial}{\partial \tilde{x}}\left(\tilde{C}_g\frac{\partial \tilde{\phi}}{\partial \tilde{x}}\right),
\label{eqn:species_transport_dim}
\end{equation}
and
\begin{equation}
    \frac{\partial \tilde{\phi}}{\partial \tilde{x}} = -\frac{-\mathcal{J} + \sum\limits_g z_g \mathscr{D}_g\frac{\partial \tilde{C}_g}{\partial \tilde{x}}}{\sum\limits_{g}z_g^2\mathscr{D}_g\tilde{C}_g},
\label{eqn:solve_for_dphi_dx_dim}
\end{equation}
where $ \tau = \frac{tD}{\ell^2} $, $\tilde{C}_g = C_g/C_0$, $\tilde{x} = x/\ell$, $\mathscr{D}_{g} = D_g / D$, $\tilde{\phi} = \phi F/(RT)$, and $\mathcal{J} = \nu\ell/(FDC_0)$. We then use eqn. \ref{eqn:copper_redox_diffusivity} to write $[\mathscr{D}_{\textrm{Cu(II)}},\mathscr{D}_{\textrm{Cu(I)}},\mathscr{D}_{\textrm{Cl$^-$}}] = [\frac{1}{2},1,1]$. We write dimensionless initial conditions as $\tilde{C}_{\textrm{Cu(II)}}(\tilde{x},\tau = 0)) = 1$, $\tilde{C}_{\textrm{Cu(I)}}(\tilde{x},\tau = 0) = 1$, and $\tilde{C}_{\textrm{Cl$^{-}$}}(\tilde{x},\tau = 0) = 3$, and boundary conditions as $\tilde{N}_{\textrm{Cu(II)}}(\tilde{x} = \pm 1) = -\mathcal{J}$, $\tilde{N}_{\textrm{Cu(I)}}(\tilde{x} = \pm 1) = \mathcal{J}$, and $\tilde{N}_{\textrm{Cl$^{-}$}}(\tilde{x} = \pm 1) = 0$, where $\tilde{N}_{g} = N_{g}\ell/(DC_0)$.

\subsubsection{Steady State Analysis}

We begin our analysis by comparing the steady-state solution of eqn. \ref{eqn:species_transport_dim} and eqn. \ref{eqn:solve_for_dphi_dx_dim} to those produced by the BSA approach. At steady-state, eqn. \ref{eqn:species_transport_dim} is equal to 0 and therefore can be written as 
\begin{equation}
    \frac{d}{d x}\tilde{N}_{g} = \frac{d}{d x}\left( -\mathscr{D}_g\frac{d\tilde{C}_g}{d \tilde{x}} - z_g\mathscr{D}_g\tilde{C}_g\frac{d \tilde{\phi}}{d \tilde{x}}\right) = 0,
\label{eqn:const_flux_nernst_1}
\end{equation}
which implies that 
\begin{equation}
    \tilde{N}_{g} =  \underbrace{-\mathscr{D}_g\frac{d\tilde{C}_g}{d \tilde{x}}}_{N_{d,g}-\textrm{diffusion}} \underbrace{- z_g\mathscr{D}_g\tilde{C}_g\frac{d \tilde{\phi}}{d \tilde{x}}}_{N_{m,g}-\textrm{electromigration}} = \textrm{constant}.
\label{eqn:const_flux_nernst}
\end{equation}
This constant is determined by evaluating the flux at one of the boundaries for each species, and therefore is given by the imposed reaction rate. In other words, we know that at steady state each total species flux must be constant throughout the electrochemical cell, or is divergence free. As we saw earlier, the BSA approach violates this in the bulk of the cell. In principle, eqn. \ref{eqn:const_flux_nernst} could be solved with eqn. \ref{eqn:solve_for_dphi_dx_dim} to determine the concentration profiles and then the relative contributions of the electromigrative and diffusive fluxes to the overall current. Ultimately, such an approach would result in an equation where $\frac{d\tilde{C}_g}{d \tilde{x}}$ would not be separable and would be under-specified due to a first-order ODE being defined by two flux boundary conditions.

For our steady-state solution we take a slightly different approach. We begin by multiplying eqn \ref{eqn:const_flux_nernst}  by $\frac{z_g}{D_g}$, summing over all ions, and applying electroneutrality to derive an alternative definition of the potential gradient as

\begin{equation}
    \frac{d\tilde{\phi}}{d\tilde{x}}=\frac{-\sum{z_g}\frac{\tilde{N}_g}{\mathscr{D}_g}}{\sum z_\textrm{g}^2\tilde{C}_{\textrm{g}}}.
    \label{eqn:potenial_simp}
\end{equation}
By inserting eqn. \ref{eqn:potenial_simp} into eqn. \ref{eqn:const_flux_nernst} we see that
\begin{equation}
    \frac{\tilde{N}_g}{\mathscr{D}_g} = -\frac{d \tilde{C}_g}{d\tilde{x}} + z_g\tilde{C}_g\frac{\sum{z_g}\frac{\tilde{C}_g}{\mathscr{D}_g}}{\sum z_g^2\tilde{C}_g}.
    \label{eqn:flux_conc_only}
\end{equation}

Now $\frac{d \tilde{C}_g}{d\tilde{x}}$ can be separated and solved using a standard boundary value problem solver. However, this version of the problem is still under-specified due to the two flux boundary conditions. A new boundary condition, i.e.,  conservation of mass for each species at steady state being equal to the initial mass, corrects the under-specification. To this end, we define $b_g$ as
\begin{equation}
    b_g = \int_{-1}^{\tilde{x}} (\tilde{C}_g-\tilde{C}_{g,0})d\tilde{x},
    \label{equ:def_of_b}
\end{equation}
where $\tilde{C}_{g,0} = \tilde{C}_{g}(\tilde{x},\tau = 0)$. With this definition, we know that $b_g(\pm 1) = 0$. We can then write eqn. \ref{equ:def_of_b} as $\frac{\partial{b_g}}{\partial{\tilde{x}}} + \tilde{C}_{g,0} = \tilde{C}_g$. We insert this into eqn. \ref{eqn:flux_conc_only} and write
\begin{equation}
    \frac{\tilde{N}_g}{\mathscr{D}_g} = z_g\left(\frac{\partial{b_g}}{\partial{\tilde{x}}} + \tilde{C}_{g0}\right) \left[ \frac{\sum_g z_g\frac{\tilde{N}_g}{\mathscr{D}_g}}{\sum_g z_g^2(\frac{\partial{b_g}}{\partial{\tilde{x}}} + \tilde{C}_{g0})}\right] - \frac{\partial^2{b_g}}{\partial{\tilde{x}^2}} 
 \label{eqn:flux_final_SS}
\end{equation}
which only depends on the parameter $\mathcal{J}$. We solve eqn. \ref{eqn:flux_final_SS} using the boundary conditions defined via eqn. \ref{equ:def_of_b}, with \texttt{Solve\_BVP} implemented in the \texttt{Scipy} library in \texttt{Python}. Afterwards, we calculate $\tilde{C}_g$ using the definition of $b_g$. Next, we discuss the steady-state concentration and flux profiles for all species as determined using the above approach.

\subsubsection{Steady-State Results}
The solution of eqn. \ref{eqn:flux_final_SS} depends on the choice of dimensionless current density $\mathcal{J}$. Ultimately, we seek to match our results to the arbitrary $k=6$ current as defined in BFW. We note that our notation for dimensionless charge flux, $\mathcal{J}$, is related to the turnover rate $k$ via
\begin{equation}
    \mathcal{J} = \frac{ke\ell}{A_\textrm{electrode}FDC_0} = \frac{k}{\gamma}.
    \label{eqn:current_density_def}
\end{equation}
This enables us to use free parameter $\gamma$, which has units of inverse time and is primarily controlled by the electrode area (which is unspecified in BFW). $\gamma$ allows us compare a simulation for a particular $\mathcal{J}$ directly to the prediction of the BSA. For this simulation, we choose $\mathcal{J} = 1/6$ which would correspond to $\gamma = 36$ [1/time]. $\mathcal{J} = 1/6$ was chosen because it corresponds to a solution that is not mass transfer limited, as discussed later. 

\begin{figure}[!ht]
\centering
\includegraphics[width=6.5in]{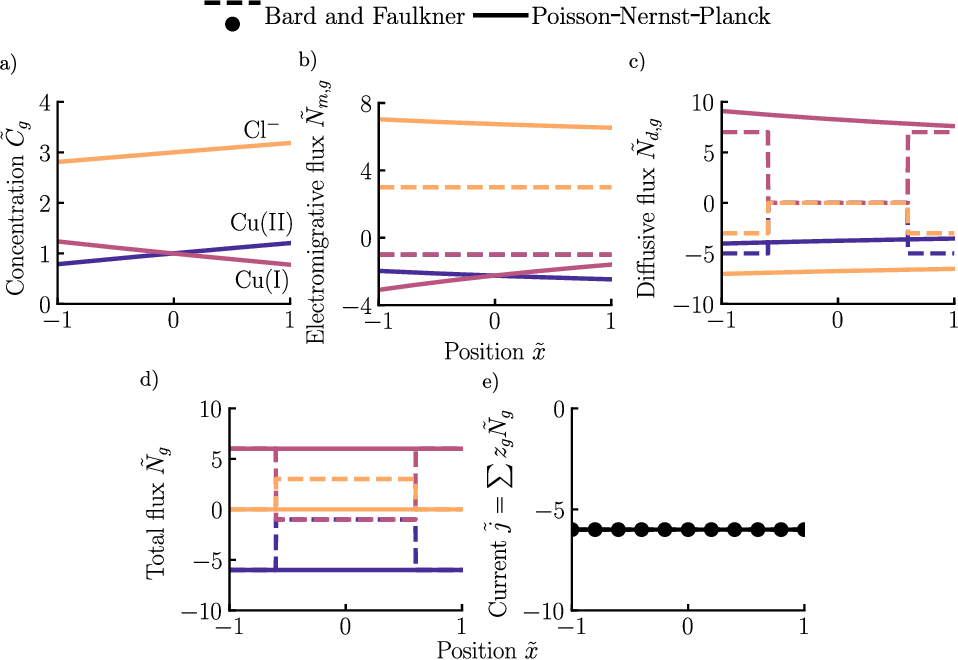}
\caption{\textbf{Concentration and flux profiles of Cu(I), Cu(II), and Cl$^-$ for the steady state copper redox cell.} The a) concentration profiles, b) electromigrative flux, and c) diffusive flux for each ionic species. d) Total flux and e) current with $\mathcal{J} = 1/6$, $\gamma = 36$, and the balance sheet approach (dashed or dotted line) for $k = 6$.}
 \label{fig3}
\end{figure}

We begin by examining the concentration profiles of Cu(II), Cu(I) and Cl$^-$ for $\mathcal{J} = 1/6$ (see Fig. \ref{fig3}a). In contrast to the BSA approach, solving the PNP equations allows one to predict the concentration profile. In Fig. \ref{fig3}a we see that the concentration of Cu(II) has a positive slope and therefore increases from left to right. This is in agreement with the consumption of Cu(II) at the cathode (left) and production at the anode (right). Cu(I) is instead produced at the cathode and consumed at the anode, it has a negative slope. Further, as Cu(II) has a larger valence $(+2)$ compared to Cu(I) $(+1)$, it therefore needs to be balanced by a larger concentration of Cl$^-$ than Cu(I) to enforce electroneutrality. This causes the concentration profile of Cl$^-$ to have a positive slope as well.

The concentration gradients are not zero anywhere in the cell, in contrast to the BSA approach, which assumes zero concentration gradients in the `bulk'. The concentration profile results are used to calculate the electromigrative and diffusive contributions to total current. The diffusive $N_{d,g}$ and electromigrative $N_{m,g}$ contributions can be calculated from total flux of each species $N_{g}$ via eqn. \ref{eqn:const_flux_nernst} where $d\tilde{\phi}/d\tilde{x}$ is determined via eqn. \ref{eqn:potenial_simp}. Fig. \ref{fig3}b and \ref{fig3}c show $N_{m,g}$ and $N_{d,g}$ respectively, which were determined by the solution of eqn. \ref{eqn:flux_final_SS} and scaled by $\gamma$ to enable a one-to-one comparison with the BSA predictions.
 
The electromigrative flux is primarily determined by the electric field generated across the cell. In agreement with the BSA approach, the electromigrative flux for both positively charged species is negative and is positive for the negatively charged species. This is driven by the positive potential gradient in the system, as the potential increases from cathode to anode. In contrast to the BSA, the electromigrative flux is not constant throughout the cell as seen in Fig. \ref{fig3}b. Instead the electromigrative flux plots are sloped and the values significantly differ from those predicted by the BSA. 

The diffusive flux in our system, as seen in Fig. \ref{fig3}c, then ensures that the total flux of each species remains constant through the cell. This requires that the Cl$^-$ diffusive flux is equal and opposite to the electromigrative Cl$^-$ flux, which yields a net Cl$^-$ flux of 0. The total flux of Cu(I) is known to be positive, so to counteract the negative electromigrative flux, the Cu(I) diffusive flux is large and positive. As Cu(II) has an overall negative flux due to its consumption at the cathode, the diffusive flux for Cu(II) is negative and smaller in magnitude than the Cu(I) diffusive flux as the electromigrative flux works in tandem with the diffusive flux. In contrast to the BSA, the diffusive fluxes calculated via solution of eq. \ref{eqn:const_flux_nernst} are not discontinuous between the bulk and the electrode. For a steady state cell, the diffusive flux is not limited to a small distance near the electrodes, as the BSA may suggest. Our approach to solving the PNP equations allows the solution to satisfy the constraint that both the total current and total flux are divergence free, i.e., they are constant throughout the entire domain as determined in eqn. \ref{eqn:constant_current} (see Fig. \ref{fig3}d and \ref{fig3}e). In comparison, the BSA only satisfies a divergence free current while the total flux divergence free condition is broken in the bulk due to the lack of diffusion.
\begin{figure}[!ht]
\centering
\includegraphics[width=6in]{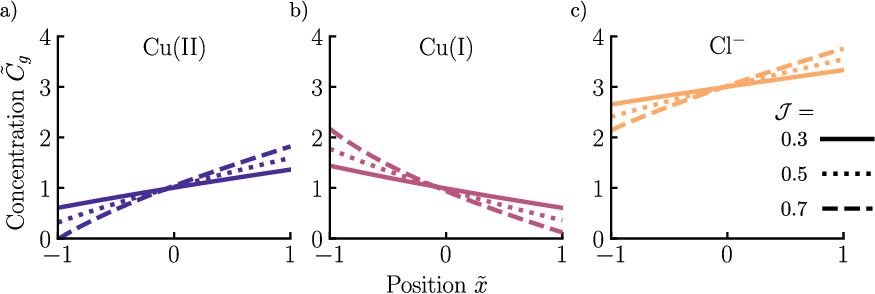}
\caption{\textbf{Concentration profiles with increasing $\mathcal{J}$ for Cu(I), Cu(II), Cl$^-$.} The concentration profiles of a) Cu(I), b) Cu(II), and c) Cl$^-$ for $\mathcal{J} = $ 0.3, 0.5, and 0.7 using the PNP simulations. The concentration of Cu(II) reaches zero at $\tilde{x} = -1$ for $\mathcal{J} = $ 0.7.}
 \label{fig4}
\end{figure}
\subsubsection{Mass Transfer Limitations at the Limiting Current}
An advantage to predicting concentration profiles is that it allows us to see mass transfer limited regimes unlike the BSA. For a zeroth-order reaction, as is used in these examples, this occurs when one of the reacting species concentrations reaches zero. The current value where mass transfer limitations occur is referred to as the limiting current \cite{selman_mass-transfer_1978}.

We determine the limiting current by increasing $\mathcal{J}$ until the concentration of one species reaches 0. For the copper redox cell, this occurs at $\mathcal{J} \approx 0.7$ where the concentration of Cu(II) drops to 0 at the cathode (see Fig. \ref{fig4}a). We note that the slopes of the concentration profiles increase in magnitude as $\mathcal{J}$ is increased.

\begin{figure}[!ht]
\centering
\includegraphics[width=5in]{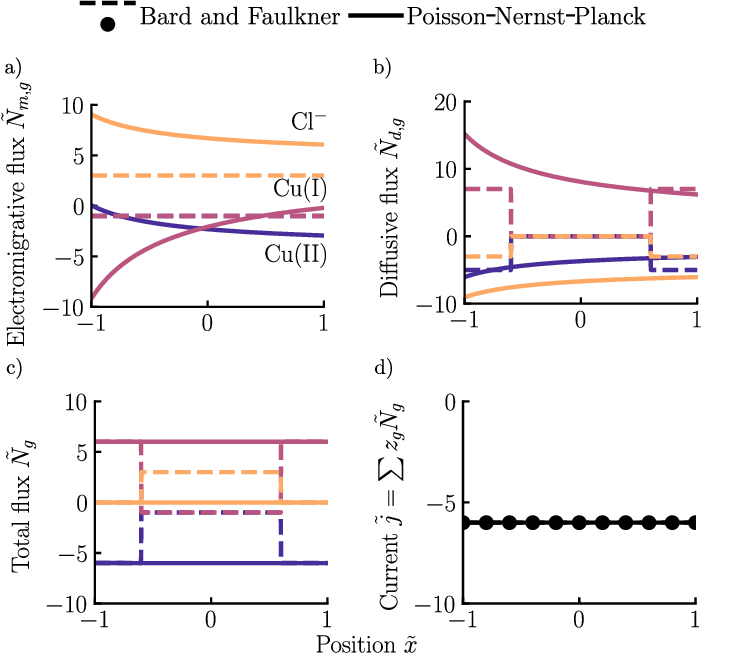}
\caption{\textbf{Flux and current profiles at $\mathcal{J} = 0.7$ for Cu(I), Cu(II), Cl$^-$}. The a) electromigrative, b) diffusive, and c) total flux for each ionic species and d) current from the solution of eqn. \ref{eqn:flux_final_SS} (solid line) for $\mathcal{J} = 0.7$, $\gamma \approx 8.7$, and the balance sheet approach (dashed or dotted line) for $k=6$.}
 \label{fig5}
\end{figure}

Fig. \ref{fig5} shows the electromigrative, diffusive, and total fluxes for each species, along with the current in the electrochemical cell for $\mathcal{J} = 0.7$ as predicted via the solution of eqn. \ref{eqn:flux_final_SS} and via the BSA approach. To reach $k=6$ with a current density of $\mathcal{J} = 0.7$, the appropriate scaling of  $\gamma \approx 8.7$ was used. Similar to the analysis for $\mathcal{J} = 1/6$, we see that the solution of the PNP equations satisfies both flux and current conservation throughout the entire domain, while the balance sheet approach only satisfies current conservation. Additionally, we see a non-zero diffusive flux in the bulk of the cell as well as a non-constant electromigrative flux from the PNP simulations, in contrast to the BSA approach.

Notably, the spatial variation, in addition to the magnitude, of the electromigrative and diffusive fluxes as determined by the PNP equations changes as $\mathcal{J}$ varies. The variation can be qualitatively seen by comparing Fig. \ref{fig5}a and \ref{fig5}b to Fig. \ref{fig3}a and \ref{fig3}b and  is not captured by the BSA approach. The arbitrary current given adds confusion to this problem since, without a given area, the given current can fall into either a reaction or mass transfer limited regime (see Figs. \ref{fig3} and \ref{fig5}). This free variable is captured by the variable $\gamma$ as defined in eqn. \ref{eqn:current_density_def}.
This analysis only applies to a system with a constant and finite reaction rate. Additionally, in a physical electrochemical cell, additional phenomena such as the reaction kinetics and the physics of the Stern and diffuse layers will influence species transport \cite{chu_electrochemical_2005} which could change the outcome of the above flux profiles.

\subsection{Transient Analysis}
We now evaluate the transient behavior of the same copper redox cell described at the beginning of section 2. At early times, i.e. before the concentration profiles in the bulk of the cell have been perturbed by the surface reaction at the electrodes, the assumption of zero bulk concentration gradients, as assumed by the BSA approach may be valid. 
Therefore, we seek to determine if there are timescales where the BSA approach accurately determines the species flux profiles. We directly solve eqn. \ref{eqn:species_transport_dim} with eqn. \ref{eqn:solve_for_dphi_dx_dim} and the initial conditions $\tilde{C}_{\textrm{Cu(II)}}(\tilde{x},\tau = 0)) = 1$, $\tilde{C}_{\textrm{Cu(I)}}(\tilde{x},\tau = 0) = 1$, and $\tilde{C}_{\textrm{Cl$^{-}$}}(\tilde{x},\tau = 0) = 3$, and boundary conditions $\tilde{N}_{\textrm{Cu(II)}}(\tilde{x} = \pm 1) = -\mathcal{J}$, $\tilde{N}_{\textrm{Cu(I)}}(\tilde{x} = \pm 1) = \mathcal{J}$, and $\tilde{N}_{\textrm{Cl$^{-}$}}(\tilde{x} = \pm 1) = 0$, using \texttt{pdepe} implemented in \texttt{MATLAB}.

\begin{figure}[!ht]
\centering
\includegraphics[width=6.5in]{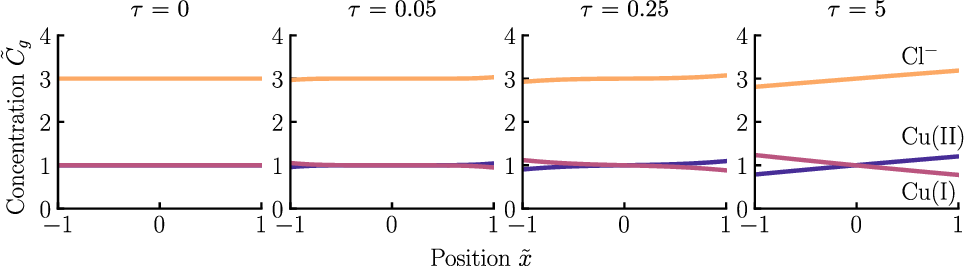}
\caption{\textbf{Transient evolution of the concentration profiles of Cu(I), Cu(II) and 
Cl$^-$} Concentration profiles of Cu(I), Cu(II) and 
Cl$^-$ for $\mathcal{J} = 1/6$ at $\tau = 0,0.05,0.25,$ and $5$ using the PNP equations.}
\label{fig6}
\end{figure}

\begin{figure}[!ht]
\centering
\includegraphics[width=5in]{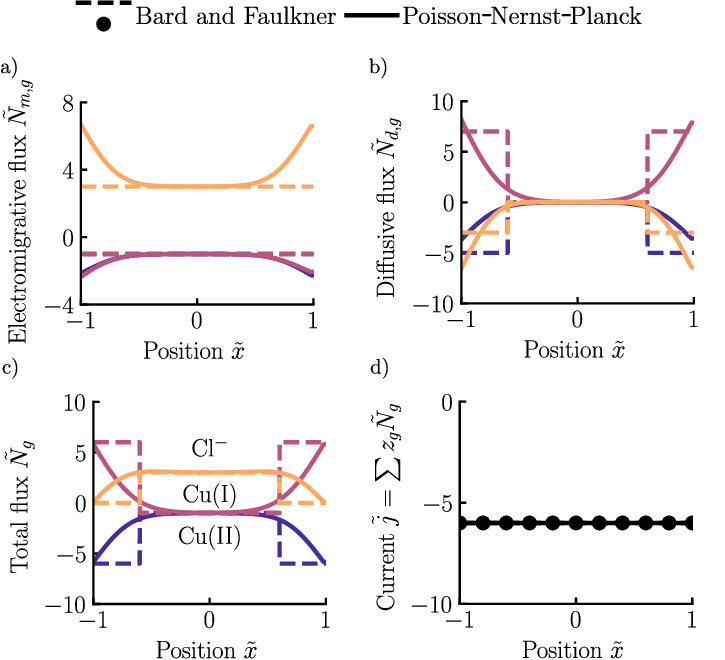}
\caption{\textbf{Transient flux and current profiles at $\tau = 0.05$ for Cu(I), Cu(II), Cl$^-$.} The a) electromigrative flux and b) diffusive flux for each ionic species, c) total flux and d) current, as determined from  the solution of eqn. \ref{eqn:species_transport_dim} (solid line) for $\mathcal{J} = 1/6$ at $\tau = 0.05$ multiplied by $\gamma = 36$ and via the ``balance sheet" approach (dashed or dotted line) for $k = \mathcal{J}\times\gamma = 6 $ electrons per unit time. }
 \label{fig7}
\end{figure}

We first examine the transient evolution of the concentration profiles for Cu(I), Cu(II) and 
Cl$^-$ as determined by solving eqn. \ref{eqn:species_transport_dim} for $\mathcal{J} = 1/6$; see Fig. \ref{fig6}. We note that by $\tau = 5$, the concentration profiles have developed to the steady-state profiles as presented in Fig. \ref{fig3}a. Additionally, at $\tau = 0.05$, we notice that the concentration profiles are relatively flat in the middle of the cell, indicating near-zero concentration gradients in that region of the cell. By $\tau= 0.25$, the concentration gradients in the bulk begin to develop. Therefore, at $\tau = 0.05$, the assumption of zero bulk concentration gradients may be valid. We used this time to evaulate electromigrative, diffusive, and total flux as well as the current  to determine if the BSA approach more accurately predicts the species contribution to flux at early times (see Fig. \ref{fig7}).

At $\tau = 0.05$, we oBSAerve quantitative agreement between the flux predictions from the PNP equations and the fluxes as predicted by the BSA approach near $\tilde{x} = 0$, i.e. in the bulk (see Fig. \ref{fig7}). This is because the zero bulk concentration gradients assumption is valid at this timescale. We see stronger deviations in the bulk as time increases to $\tau = 0.25$ and eventually approaches steady state, as previously discussed with Fig. \ref{fig3}. Near both electrodes, i.e. at $\tilde{x} = \pm 1$, we see a strong deviation between the prediction from the PNP and BSA approaches. The deviation indicates that it was incorrect to assume that the electromigrative flux throughout the cell remains spatially constant at the value determined in the bulk of the cell. The bulk electromigrative flux, as determined via the BSA, was calculated using the transference number definition that only holds with zero concentration gradients (see eqn. \ref{eqn:transference_number}). As shown in Fig. \ref{fig6}, this is not the case near the electrodes, even at $\tau = 0.05$.

Furthermore, we see that the total species flux is no longer conserved, as transient accumulation can occur. We note that one benefit of the PNP equations is that it predicts a smooth, rather than discontinuous, transition from the bulk flux values to those near the electrodes. Lastly, we see that the current in the cell remains constant in both space and time, as predicted by Eqn. \ref{eqn:constant_current}.

\subsection{Mixing Analysis}
\begin{figure}[!ht]
\centering
\includegraphics[width=5.26in]{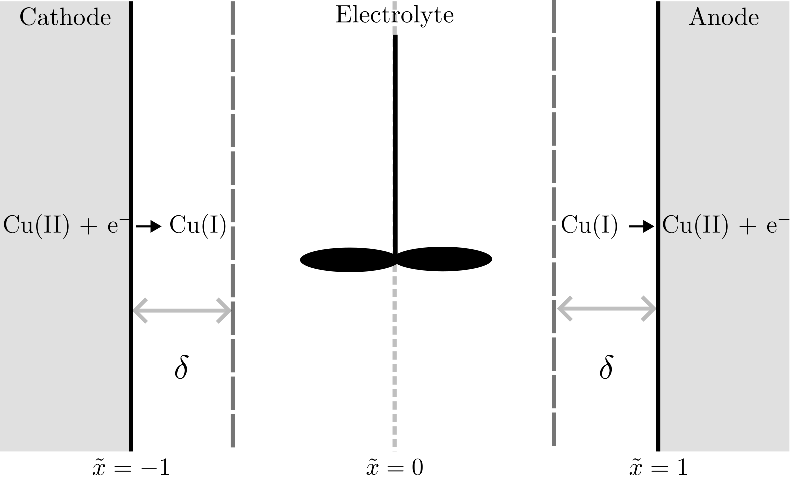}
\caption{\textbf{Schematic of mixing in a copper redox cell.} An electrolyte solution composed of $10^{-3}$ M Cu(NH$_3)_4^{2+}$, $10^{-3}$ M Cu(NH$_3)_2^{+}$, and $3 \times 10^{-3}$ M Cl$^-$ in $0.1$ M NH$_3$ is placed in an electrochemical cell with a cathode at $x = -\ell$ and an anode at $x = \ell$. A current is produced via the reduction of Cu(II) at  the cathode and the oxidation of Cu(I) at the electrode. Constant mixing due to a impeller occurs in the center of the cell generating diffusion layers with constant thickness $\delta$.}
 \label{fig8}
\end{figure}

Another mechanism through which the assumption of zero concentration gradients could be attained in the bulk is through mixing. Adding agitation would cause all concentration gradients to be uniform. This idea comes from a source referenced in BFW called Electrochemical Reactions by G. Charlot, J. Badoz-Lambling, and B. Tr\'emillon \cite{charlot_electrochemical_1962}. In Electrochemical Reactions, the same balance sheet setup is used as in Section 2.1 but mixing in the bulk is included and used as an explanation for constant concentration gradients. In the following we describe our formulation of a solution to the problem using the PNP equations for a copper redox cell with constant bulk mixing. We assume that a bulk region from ($-1 +\delta$) to ($1-\delta$) is being constantly stirred such that the bulk region is well mixed and therefore the concentration profiles for all species do not have spatial dependence. We assume that in a small region near both electrodes, with a thickness $\delta$, the fluid velocity is small enough that diffusion and electromigration dominate over convection \cite{charlot_electrochemical_1962}. We will call this region a so-called ``diffusion layer." A schematic illustration of this setup can be see in Fig. \ref{fig8}, where the mixing is indicated by an impeller. We then piecewise define $\tilde{N}_{g}$ and $\tilde{C}_{g}$ in the diffusion layer near the cathode, in the bulk, and in the diffusion layer near the anode via

\begin{equation}
    \tilde{C}_{g} = \begin{cases} 
      \tilde{C}_{\textrm{cathode},g} & -1\leq \tilde{x}\leq -1 + \delta \\
      \tilde{C}_{\textrm{bulk},g} & -1 + \delta < \tilde{x} < 1-\delta \\
      \tilde{C}_{\textrm{anode},g} & 1-\delta\leq \tilde{x}\leq 1
   \end{cases}
   \label{eqn:mixing_conc}
\end{equation}
and
\begin{equation}
    \tilde{N}_{g} = \begin{cases} 
      \tilde{N}_{\textrm{cathode},g} & -1\leq \tilde{x}\leq -1 + \delta \\
      \tilde{N}_{\textrm{bulk},g} & -1 + \delta < \tilde{x} < 1-\delta \\
      \tilde{N}_{\textrm{anode},g} & 1-\delta\leq \tilde{x}\leq 1.
   \end{cases}
   \label{eqn:mixing_flux}
\end{equation}

In the diffusion layers near the cathode and anode we solve the steady-state PNP equations as described in section 2.2.3 for each region separately. At the cathode we use the flux conditions $\tilde{N}_{\textrm{cathode,Cu(II)}}(\tilde{x} = - 1) = -\mathcal{J}$, $\tilde{N}_{\textrm{cathode,Cu(I)}}(\tilde{x} = - 1) = \mathcal{J}$, and $\tilde{N}_{\textrm{cathode,Cl$^{-}$}}(\tilde{x} = - 1) = 0$. We close the problem by assuming that $\tilde{C}_{\textrm{cathode},g}(\tilde{x} = -1 + \delta) = \tilde{C}_{\textrm{bulk},g}(\tilde{x} = -1 + \delta)$. In the diffusion layer near the anode, we use the boundary conditions $\tilde{N}_{\textrm{anode,Cu(II)}}(\tilde{x} = + 1) = -\mathcal{J}$, $\tilde{N}_{\textrm{anode,Cu(I)}}(\tilde{x} = + 1) = \mathcal{J}$, $\tilde{N}_{\textrm{anode,Cl$^{-}$}}(\tilde{x} = + 1) = 0$, and $\tilde{C}_{\textrm{anode},g}(\tilde{x} = 1- \delta) = \tilde{C}_{\textrm{bulk},g}(\tilde{x} = 1- \delta)$. 

To determine the bulk concentration, we write a mole balance for each species in the bulk as
\begin{equation}
   V_{\textrm{bulk}}\frac{d\tilde{C}_{\textrm{bulk,g}}}{d\tau} = \tilde{N}_{\textrm{bulk,g}}(\tilde{x} = -1 + \delta)A_{\textrm{electrode}} - \tilde{N}_{\textrm{bulk,g}}(\tilde{x} = 1- \delta)A_{\textrm{electrode}}.
\end{equation}
Using the steady-state assumption, $\frac{d\tilde{C}_{\textrm{bulk,g}}}{d\tau} = 0$ which implies that $\tilde{C}_{\textrm{bulk,g}} = \textrm{constant}$. Evaluating at the initial conditions we write that $\tilde{C}_{\textrm{bulk,g}}  = \tilde{C}_{g}(\tau = 0)$. Hence, the concentrations in the bulk (and at the interface between the diffusion layers and the bulk) remain spatially and temporally constant at the values defined by the initial conditions. We note that this also implies that $\tilde{N}_{\textrm{bulk,g}}(\tilde{x} = -1 + \delta) = \tilde{N}_{\textrm{bulk,g}}(\tilde{x} = 1- \delta)$. This analysis does not provide information about which mechanisms, i.e. electromigration, convection, or diffusion, are responsible for carrying species flux, and therefore the current, throughout the bulk. Such a determination would require a direct numerical simulation which is outside the scope of this work. Along the same lines, we note that although the BSA analysis can be interpreted as supporting zero bulk concentration gradients through mixing, the mechanisms proposed for bulk current transport are not internally consistent.

\subsection{Results}
\begin{figure}[!ht]
\centering
\includegraphics[width=5in]{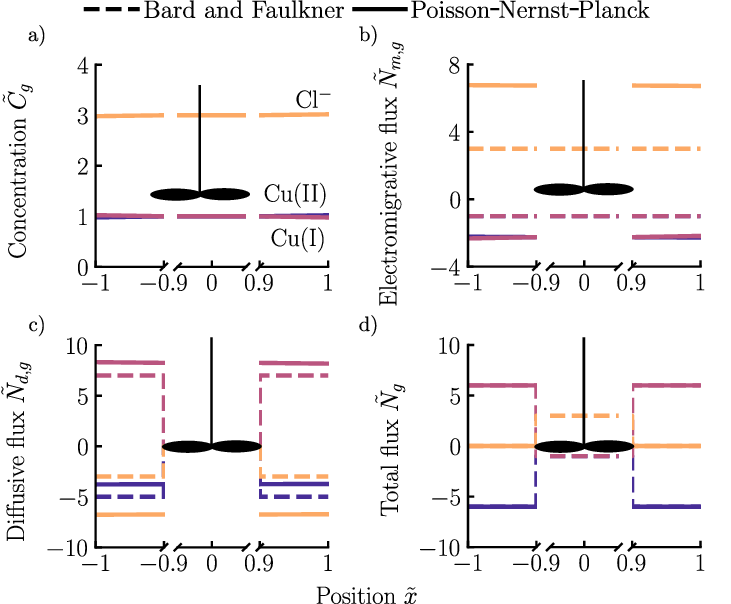}
\caption{\textbf{Mixing flux and concentration profiles for  Cu(I), Cu(II), Cl$^-$.} The a) concentration, b) electromigraive flux, c) diffusive flux, and d) total flux profiles for each ionic species, as determined from the solution of eqn. \ref{eqn:const_flux_nernst} (solid line) for $\mathcal{J}=1/6$ multiplied by $\gamma=36$, along with the balance sheet approach (dashed line) for $k=\mathcal{J}\times\gamma=6$ electrons per unit time.}
 \label{fig9}
\end{figure}

We begin by plotting the flux and concentration profiles, as piecewise defined in eqns. \ref{eqn:mixing_conc} and \ref{eqn:mixing_flux}, and solved for as described above, for an arbitrary $\delta = 0.1$, $\mathcal{J}= 1/6$, and $\gamma = 36$; see Fig. \ref{fig9}. A more physical route for the determination of $\delta$ would require knowledge of the cell geometry and the so-called ``mixing Reynolds number." We expect $\delta = 0.1 \ll 1$ to be representative of a typical system, but acknowledge that it was chosen arbitrarily. We believe this to be justifiable given that our goal is pedagogical in nature.

We oBSAerve that the concentration profiles vary slightly in diffusive regions, see Fig. \ref{fig9}a. In Fig. \ref{fig9}b/c, we oBSAerve that the direction of the diffusive and electromigrative fluxes agree with the predictions from the BSA \cite{bard2022electrochemical}, as was the case in the previous transient and steady state analysis. However, we see quantitative disagreement between the magnitudes of the flux values. Similar to the steady-state analysis, the total flux near the electrodes, matches the imposed reaction rate at the boundaries. In the bulk regions we are unable to compare our flux calculations to those of the BSA as full CFD calculations are required to resolve the fluxes accurately.

\section{Copper Redox Cell with Supporting Electrolyte}
\begin{figure}[!ht]
\centering
\includegraphics[width=5.26in]{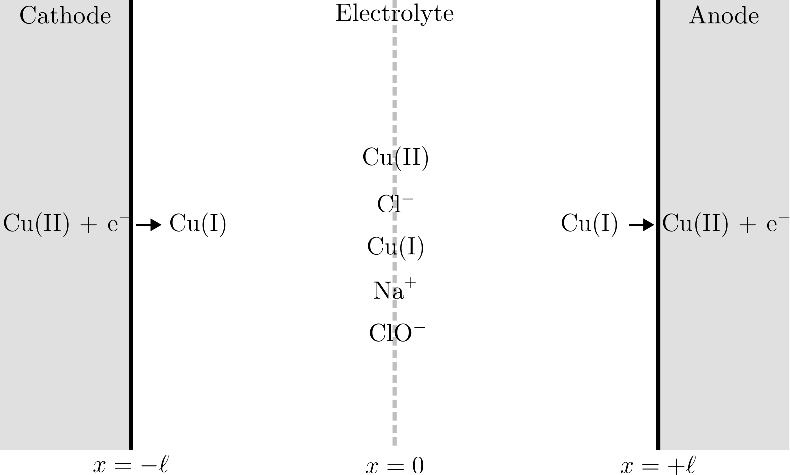}
\caption{\textbf{Schematic representation of copper redox cell with supporting electrolyte presented from Section 4.3 in Bard, Faulkner and White}. An electrolyte solution with $10^{-3}$ M Cu(NH$_3)_4^{2+}$, $10^{-3}$ M Cu(NH$_3)_2^{+}$, $3\times10^{-3}$ M Cl$^-$ in $0.1$ M NH$_3$, and NaClO$_4$ at $0.1$ M are placed in an electrochemical cell with a cathode at $x=-\ell$ and an anode at $x=\ell$.}
 \label{fig10}
\end{figure}

After comparing the concentration profiles and flux predictions for the copper redox cell, as determined via solution of the PNP equations and via the BSA approach, we continue our analysis by looking at the same copper redox cell in the previous example but with the addition of a supporting electrolyte. This follows Figure 4.3.4 in BFW \cite{bard2022electrochemical}. This one-dimensional electrochemical cell still has $10^{-3}$ M Cu(NH$_3)_4^{2+}$, $10^{-3}$ M Cu(NH$_3)_2^{+}$, and $3\times10^{-3}$ M Cl$^-$ in $0.1$ M NH$_3$, but now has added NaClO$_4$ at $0.1$ M, a concentration 100 times higher than each reacting copper ion. The Na$^+$ and ClO$_4^-$ do not participate in the redox reactions at either electrode. The setup is schematically depicted in Fig. \ref{fig10}. 

The equivalent conductance ($\lambda_g$) for all species is assumed equal, i.e. $\lambda_{\textrm{Cu(II)}} = \lambda_{\textrm{Cu(I)}} = \lambda_{\textrm{Cl$^-$}} = \lambda_{\textrm{Na$^+$}} = \lambda_{\textrm{ClO$_4^-$}} = \lambda$. We then use eqn. \ref{eqn:lim_cond} the diffusivity of each species as
\begin{equation}
\begin{matrix}      D_{\textrm{Cu(II)}}=\\ 
    D_{\textrm{Cu(I)}} =\\
    D_{\textrm{Cl}^{-}}=\\
    D_{\textrm{Na}^{+}}=\\
    D_{\textrm{ClO$_4^-$}} =
\end{matrix}  
\begin{matrix}
    \frac{D}{2}, \\ D, \\ D \\ D, \\ D.
\end{matrix},
\label{eqn:SE_diff}
\end{equation}
 
\subsection{Balance Sheet Approach}
We repeat the set up from section 2.1 with the same copper redox reactions at each electrode and $k=6$ with 6 Cu(II) atoms per unit time being consumed and 6 Cu(I) atoms per unit time being produced. The inverse occurs at the anode. The key difference here is the addition of NaClO$_4$ in the solution. Cl$^-$, Na$^+$, and ClO$_4^-$ do not participate in the reactions. 

Assuming no bulk concentration gradients, transference numbers for each ion were calculated using eqn. \ref{eqn:transference_number} so that eqn. \ref{eqn:em_current} could be used to calculate $i_{m,g}$ as
\begin{equation}
\begin{matrix}
    i_{m,\textrm{Cu(II)}} \\ 
    i_{m,\textrm{Cu(I)}} \\
    i_{m,\textrm{Cl}^{-}} \\
    i_{m,\textrm{Na}^{+}} \\
    i_{m,\textrm{ClO$_4^-$}}
\end{matrix} = 
\begin{matrix}
    -0.06 \\ -0.03 \\ 
-0.09 \\ -2.91 \\ -2.91
\end{matrix}.
\end{equation}
We can then write the atomic rate of each species in the bulk, which is assumed to be due only to electromigration, using eqn \ref{equn:atomic_current_def} as
\begin{equation}
\begin{matrix}
    n_{\textrm{Cu(II)}} \\ 
    n_{\textrm{Cu(I)}} \\
    n_{\textrm{Cl}^{-}} \\
    n_{\textrm{Na}^{+}} \\
    n_{\textrm{ClO$_4^-$}}
\end{matrix} = 
\begin{matrix}
    -0.03 \\ -0.03 \\ 0.09 \\ -2.91 \\ 2.91
\end{matrix}.
\end{equation}

The next step is to balance the atomic rates at the electrodes using diffusive contribution to equal the overall reaction rate. The electromigrative rate remains constant while the species atomic rates are balanced with diffusion even though electromigrative rate values were calculated assuming no concentration gradients which no longer holds true near the electrodes. The overall atomic rate at the cathode can be written as
\begin{equation}
\begin{matrix}
    n_{\textrm{Cu(II)}}, \\ 
    n_{\textrm{Cu(I)}}, \\
    n_{\textrm{Cl}^{-}}, \\
    n_{\textrm{Na}^{+}}, \\
    n_{\textrm{ClO$_4^-$}}.
\end{matrix} = 
\begin{matrix}
    -6 \\ 6 \\ 0 \\ 0 \\ 0
\end{matrix}.
\end{equation}

\begin{figure}[!ht]
\centering
\includegraphics[width=5.1in]{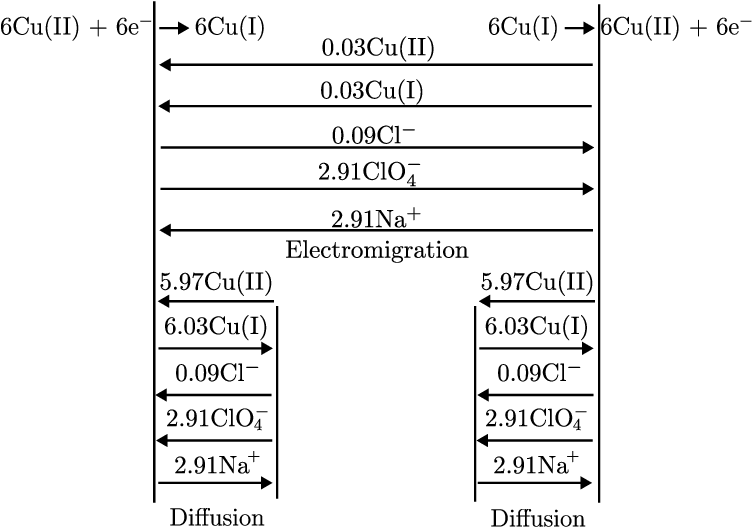}
\caption{\textbf{Recreation of Figure 4.3.4 from Bard, Faulkner and White \cite{bard2022electrochemical}.} Depiction of electromigrative and diffusive rate contributions as determined by the ``balance sheet" procedure.}
 \label{fig11}
\end{figure}

From here, diffusive rates can be determined at the cathode with the expression $n_{d,g} = n_{g} - n_{m,g}$ as
\begin{equation}
\begin{matrix}
    n_{d,\textrm{Cu(II)}}, \\ 
    n_{d,\textrm{Cu(I)}}, \\
    n_{d,\textrm{Cl}^{-}}, \\
    n_{d,\textrm{Na}^{+}}, \\
    n_{d,\textrm{ClO$_4^-$}}.
\end{matrix} = 
\begin{matrix}
    -5.97 \\ 6.03 \\ -0.09 \\ 2.91 \\ -2.91
\end{matrix}.
\end{equation}
With these values for atomic rates, a balance sheet can be formulated for this example (see Fig. \ref{fig11}).

The atomic rate values in the bulk for the reactive species, i.e. Cu(I) and Cu(II), are much smaller than the than those of the supporting electrolyte. Thus, according to BSA analysis, the supporting electrolyte carries the majority of the current. However, as noted earlier, at steady state both the current and species fluxes must be divergence free. This again begs the question of how the necessary 6 Cu(I) and Cu(II) ions are being transported through the bulk, as the BSA analysis states that only 0.03 ions per units time of Cu(I) and Cu(II) are being transported in the bulk. In the following section we compare the predictions from the solution of the steady state PNP equations to those predicted by the balance sheet.

\subsection{Analysis via PNP Equations}
Adding NaClO$_4$ as a supporting electrolyte adds two more ions to the copper redox problem and therefore two more equations, initial conditions, and boundary conditions to the problem set. We can then use eqn. \ref{eqn:SE_diff} to write $[\mathscr{D}_{\textrm{Cu(II)}},\mathscr{D}_{\textrm{Cu(I)}},\mathscr{D}_{\textrm{Cl$^-$}},\mathscr{D}_{\textrm{Na}^+}, \mathscr{D}_{\textrm{ClO}_4^-}] = [\frac{1}{2},1,1,1,1]$. Additionally, initial conditions are written as $\tilde{C}_{\textrm{Cu(II)}}(\tilde{x},\tau = 0) = 1$, $\tilde{C}_{\textrm{Cu(I)}}(\tilde{x},\tau = 0) = 1$, $\tilde{C}_{\textrm{Cl$^{-}$}}(\tilde{x},\tau = 0) = 3$, $\tilde{C}_{\textrm{Na}^+}(\tilde{x},\tau = 0) = 100$, and $\tilde{C}_{\textrm{ClO}_4^-}(\tilde{x},\tau = 0) = 100$
and boundary conditions as $\tilde{N}_{\textrm{Cu(II)}}(\tilde{x} = \pm 1) = -\mathcal{J}$, $\tilde{N}_{\textrm{Cu(I)}}(\tilde{x} = \pm 1) = \mathcal{J}$, and $\tilde{N}_{\textrm{Cl$^{-}$}}(\tilde{x} = \pm 1) = \tilde{N}_{\textrm{Na$^{+}$}}(\tilde{x} = \pm 1)= \tilde{N}_{\textrm{ClO$_4^{-}$}}(\tilde{x} = \pm 1)= 0$. We then solve \ref{eqn:flux_final_SS} with boundary conditions defined via eqn. \ref{equ:def_of_b} using \texttt{Solve\_BVP} implemented in \texttt{Scipy} library in \texttt{Python} to determine the steady-state flux and species concentration profiles.

\begin{figure}[!ht]
\centering
\includegraphics[width=5in]{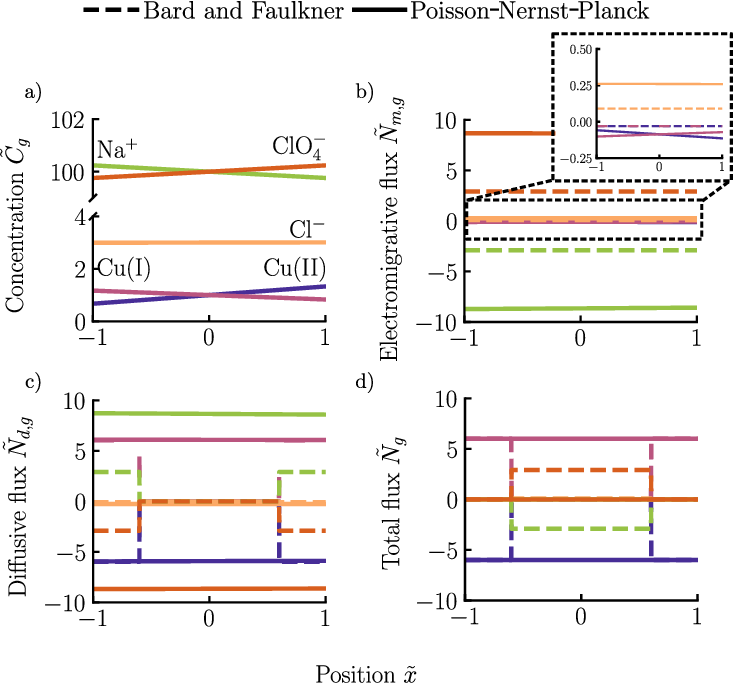}
\caption{\textbf{Concentration and flux profiles of Cu(I), Cu(II), Cl$^-$, Na$^+$, ClO$_4^-$ for the steady-state copper redox cell with a supporting electrolyte.} a) Steady-state concentration profiles for each species as determined via solution of eqn. \ref{eqn:flux_final_SS} for $\mathcal{J} = 1/6$. The b) electromigrative flux, c) diffusive flux for each ionic species, and d) total flux as determined from  the solution of eqn. \ref{eqn:flux_final_SS} (solid line) for $\mathcal{J} = 1/6$ multiplied by $\gamma = 36$ and via the balance sheet approach (dashed line) for $k = 6$ electrons per unit time.}
 \label{fig12}
\end{figure}

 We begin by looking at the concentration profiles of Cu(II), Cu(I), and Cl$^-$ as seen in Fig. \ref{fig12}a. The concentration profiles for Cu(II) and Cu(I) are similar to the profiles from Fig. \ref{fig3}a with a positive slope for Cu(II) because of consumption at the left cathode and a negative slope for Cu(I) with consumption at the right anode. Relative to Fig. \ref{fig3}a, the Cl$^-$ profile is exhibits a much smaller slope. Physically, this can be interpreted as the excess electrolyte playing a larger role in maintaining electroneutrality, reducing the necessity for Cl$^-$ to do such. We notice much larger variations when comparing the electromigrative fluxes for Cu(II), Cu(I), and Cl$^-$ between Fig. \ref{fig3} and Fig \ref{fig12}. In general, all three species exhibit much smaller electromigrative fluxes in the presence of a supporting electrolyte. Examination of eqn. \ref{eqn:solve_for_dphi_dx_dim} indicates that $d\tilde{\phi}/d\tilde{x}$ is inversely proportional to the total concentration in the electrolyte. Therefore, the large concentration of NaClO$_4$, serves the role of decreasing the potential gradient and hence the electromigrative fluxes. As Na$^+$ and ClO$_4^-$ are present in large concentrations, they exhibit electromigrative fluxes of a larger magnitude. Additionally, $\tilde{N}_{\textrm{ClO$_4^-$}}$ is positive, indicating a flux towards the anode, while $\tilde{N}_{\textrm{Na$^+$}}$ exhibits a negative flux, indicating a flux towards the cathode. This is in line with the positive potential gradient that exists in the system, as the potential increases from the cathode to the anode. Similar to Fig. \ref{fig3}, the diffusive flux (see Fig. \ref{fig12}c) then ensures that the total flux (see Fig. \ref{fig12}d) for all species remains divergence-free. When compared to the BSA approach, generally we see agreement with the direction of electromigrative and diffusive flux, however, there is significant quantitative disagreement. Furthermore, the BSA approach is still unable to provide information about concentration profiles, predicts discontinous diffusive and total flux profiles, and does not satisfy a divergence-free total flux.

\section{Hydrogen Evolution Cell, Unsteady State}
\begin{figure}[!ht]
\centering
\includegraphics[width=5.26in]{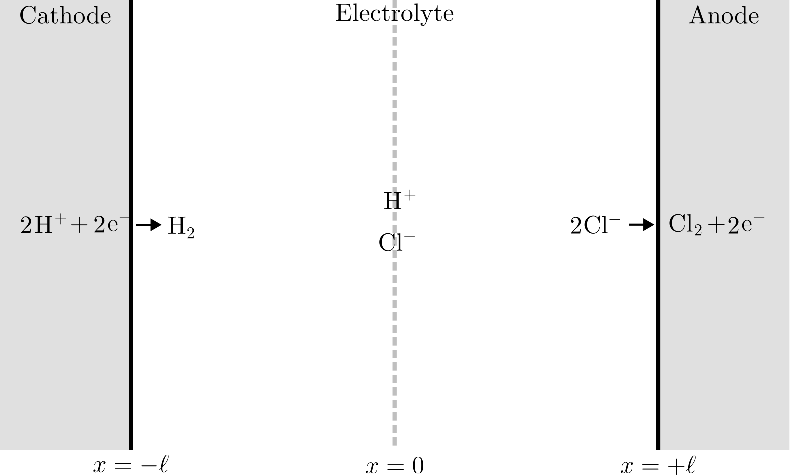}
\caption{\textbf{Schematic representation of a hydrogen evolution cell from Section 4.3 in Bard, Faulkner and White}. A salt solution of HCl is added to an electrochemical cell with a platinum cathode and anode. Upon application of a potential a current is produced via the reduction of H$^+$ at the cathode and the oxidation of Cl$^-$ at the electrode.}
 \label{fig13}
\end{figure}

The last example from BFW we will analyze is the unsteady hydrogen evolution cell as described in figure 4.3.2 in Chapter 4 Section 3. In this example, BFW  apply the BSA approach to the electrolysis of hydrochloric acid at two platinum electrodes. At the left electrode, located at $x = -l$, H$^+$ is reduced to form H$_2$ by consuming electrons. At the right electrode, located at $x = l$, Cl$^-$ is oxidized to form Cl$_2$, producing electrons in the process; see Fig. \ref{fig13}. Given that neither Cl$^-$ or H$^+$ are being replenished, the concentrations of both ionics species will deplete over time, meaning that the problem is inherently transient.

It is assumed that the H$^+$ and Cl$^-$ equivalent conductance are related via $\lambda_{H+} \approx 4\lambda_{Cl-}$. Based on eqn. \ref{eqn:lim_cond} the diffusivities are given as
\begin{equation}
\begin{matrix}
    D_{\textrm{H}} \\ 
    D_{\textrm{Cl}^{-}} \\
\end{matrix} = 
\begin{matrix}
    D, \\ \frac{D}{4}. 
\end{matrix}
\end{equation}
where $D$ is the diffusivity of H$^+$. 

\subsection{Balance Sheet Approach}
In BFW, a constant turnover rate of $k=10$ electrons per unit time is assumed. We note that in a realistic setting, the reaction rate will likely depend on the concentration of each species. The turnover rate of $k=10$ electrons per unit implies that at the cathode 10 H$^+$ atoms are consumed and 5 H$_2$ atoms are produced. At the anode 10 Cl$^-$ atoms are consumed and 5 Cl$_2$ atoms produced. Using the definition of current provided in eqn. \ref{equn:atomic_current_def} we see that, when evaluated at the left electrode, $i = z_{\textrm{H}^+}n_{\textrm{H}^+} = 1\times(-10) = -10$, which corresponds to a current of -10 electrons per unit time flowing from left to right in the cell. 

As before, it is assumed that electromigration dominates away from the electrodes. Using the transference number as defined in eqn. \ref{eqn:transference_number}, we find the electromigrative contribution to the current of each species as $i_{m,g}$ as
\begin{equation}
\begin{matrix}
    i_{m,\textrm{H}^+} \\ 
    i_{m,\textrm{Cl}^-}
\end{matrix} = 
\begin{matrix}
    -8, \\ -2.
\end{matrix}
\end{equation}
We can then use eqn. \ref{equn:atomic_current_def} to write the atomic rate of each species in the bulk, which is assumed to be due to only electromigration, as
\begin{equation}
\begin{matrix}
    n_{\textrm{H}^+} \\ 
    n_{\textrm{Cl}^-}
\end{matrix} = 
\begin{matrix}
    -8, \\ 2.
\end{matrix}
\end{equation}

The next step in the balance sheet approach is to balance species atomic rates at the anode and cathode by using the diffusive rate. The overall atomic rate at the cathode can be written as
\begin{equation}
\begin{matrix}
    n_{\textrm{H}^+} \\ 
    n_{\textrm{Cl}^-}
\end{matrix} = 
\begin{matrix}
    -10, \\ 0,
\end{matrix}
\end{equation}
and at the anode as
\begin{equation}
\begin{matrix}
    n_{\textrm{H}^+} \\ 
    n_{\textrm{Cl}^-}
\end{matrix} = 
\begin{matrix}
    0, \\ 10.
\end{matrix}
\end{equation}
We then determine diffusive with $n_{d,g} = n_g-n_{m,g}$ at the cathode as
\begin{equation}
\begin{matrix}
    n_{d,\textrm{H}^+} \\ 
    n_{d,\textrm{Cl}^-}
\end{matrix} = 
\begin{matrix}
    -2, \\ -2,
\end{matrix}
\end{equation}
and at the anode as
\begin{equation}
\begin{matrix}
    n_{d,\textrm{H}^+} \\ 
    n_{d,\textrm{Cl}^-}
\end{matrix} = 
\begin{matrix}
    8, \\ 8.
\end{matrix}
\end{equation}

\begin{figure}[!ht]
\centering
\includegraphics[width=5.1in]{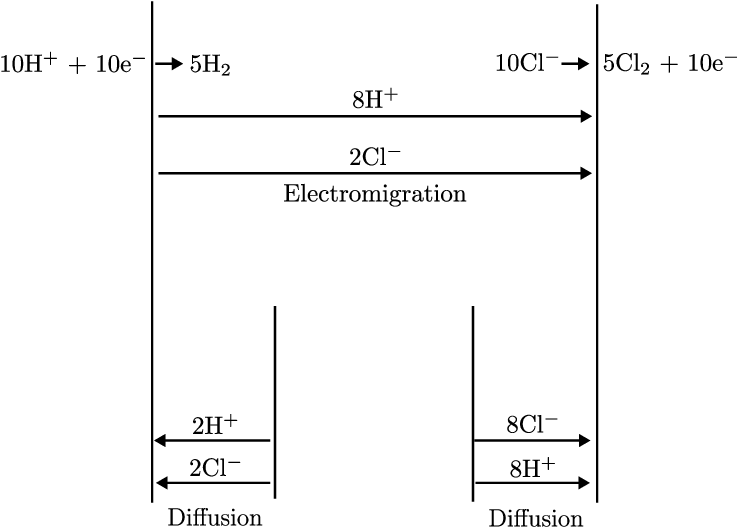}
\caption{\textbf{Recreation of Figure 4.3.2 from Bard, Faulkner and White \cite{bard2022electrochemical}.} Depiction of electromigrative and diffusive rate contributions as determined by the balance sheet  for the electrolysis of hydrochloric acid.}
 \label{fig14}
\end{figure}

In addition to the questions raised earlier regarding the BSA predictions, an important issue in this example concerns the time scale over which the BSA predictions are formulated. As is apparent, this problem is inherently transient, therefore it is expected that the flux profiles will change transiently as the concentrations of H$^+$ and Cl$^-$ are depleted. A schematic depiction of the balance sheet can be seen in Fig. \ref{fig14}.

\subsection{Transport Equation Analysis}
As in section 2.3, eqn. \ref{eqn:species_transport_dim} with eqn. \ref{eqn:solve_for_dphi_dx_dim} could be directly solved to determine the concentration and flux profiles for H$^+$ and Cl$^-$ with the appropriate boundary and initial conditions. We note that a long time solution is physically ill-posed since a continuous depletion of ions would render the concentrations negative. Given this inconsistency and given the pedagogical nature of this work, we will only focus on the early-time solutions. It is interesting to note that this problem permits a similarity variable solution which is valid for $t\ll \ell^2/D$, or physically speaking, before the depletion of the concentration of H$^+$ and Cl$^-$ reach the middle $x = 0$ of the electrochemical cell. In the following we will demonstrate how to arive at such a solution.
We begin by writing eqn. \ref{eqn:species_transport_dim} for $\textrm{H}^+$ and $\textrm{Cl}^-$ as
\begin{equation}
    \frac{\partial\tilde{C}_\textrm{H$^+$}}{\partial\tau}=\mathscr{D}_\textrm{H$^+$}\frac{\partial^2\tilde{C}_\textrm{H$^+$}}{\partial\tilde{x}^2}+\mathscr{D}_\textrm{H$^+$}\frac{\partial}{\partial\tilde{x}}\left(\tilde{C}_\textrm{H$^+$}\frac{\partial \tilde{\phi}}{\partial\tilde{x}}\right),
    \label{eqn:H_species_balance}
\end{equation}
\begin{equation}
    \frac{\partial\tilde{C}_{\textrm{Cl$^-$}}}{\partial\tau}=\mathscr{D}_{\textrm{Cl$^-$}}\frac{\partial^2\tilde{C}_{\textrm{Cl$^-$}}}{\partial\tilde{x}^2}-\mathscr{D}_{\textrm{Cl$^-$}}\frac{\partial}{\partial\tilde{x}}\left(\tilde{C}_{\textrm{Cl$^-$}}\frac{\partial \tilde{\phi}}{\partial\tilde{x}}\right).
    \label{eqn:Cl_species_balance}
\end{equation}
Using electroneutrality $z_{\textrm{Cl$^-$}}\tilde{C}_{\textrm{Cl$^-$}} + z_\textrm{H$^+$}\tilde{C}_{\textrm{H$^+$}} = 0$ and that $z_{\textrm{Cl$^-$}} = -z_\textrm{H$^+$}$ we write $\tilde{C}_{\textrm{Cl$^-$}}=\tilde{C}_{\textrm{H}} = \tilde{C_\textrm{S}}$ with $\tilde{C}_\textrm{S}$ being the salt concentration. Eqns.  \ref{eqn:H_species_balance} and \ref{eqn:Cl_species_balance} can be written using $\tilde{C}_\textrm{S}$ as
\begin{equation}
    \frac{1}{\mathscr{D}_\textrm{H}}\frac{\partial\tilde{C}_\textrm{S}}{\partial\tau}=\frac{\partial^2\tilde{C}_\textrm{S}}{\partial\tilde{x}^2}+\frac{\partial}{\partial\tilde{x}}\left(\tilde{C}_\textrm{S}\frac{\partial \tilde{\phi}}{\partial\tilde{x}}\right),
    \label{eqn:H_sb_C}
\end{equation}
\begin{equation}
    \frac{1}{\mathscr{D}_{\textrm{Cl}}}\frac{\partial\tilde{C}_\textrm{S}}{\partial\tau}=\frac{\partial^2\tilde{C}_\textrm{S}}{\partial\tilde{x}^2}-\frac{\partial}{\partial\tilde{x}}\left(\tilde{C}_\textrm{S}\frac{\partial \tilde{\phi}}{\partial\tilde{x}}\right).
    \label{eqn:Cl_sb_C}
\end{equation}
Adding eqns. \ref{eqn:H_sb_C} to \ref{eqn:Cl_sb_C} yields
\begin{equation}
    \frac{\partial\tilde{C}_\textrm{S}}{\partial\tau}=D_a\frac{\partial^2\tilde{C}_\textrm{S}}{\partial \tilde{x}^2},
    \label{eqn:salt_diffusion}
\end{equation}
where $D_a = \left(2\mathscr{D}_\textrm{H$^+$}\mathscr{D}_\textrm{Cl$^-$}\right)/\left(\mathscr{D}_\textrm{H$^+$}+\mathscr{D}_\textrm{Cl$^-$}\right)$ is the nondimensional ambipolar diffusivity. Rather than having to solve the coupled eqns. \ref{eqn:H_species_balance} and \ref{eqn:Cl_species_balance} with an appropriate mechanism for determining the potential gradient, we simply have to solve a single transient diffusion equation for the salt concentration. After solving the diffusion equation, the fluxes and concentrations of H$^+$ and Cl$^-$ can be determined. Before solving, we must determine appropriate boundary condition and initial conditions for the salt concentrations. We assume there is an initial uniform concentration of salt $\tilde{C_{\textrm{S}}}(\tilde{x},\tau =0)= 1$. At the cathode, we write that the fluxes are equivalent to the imposed surface reaction rate $\mathcal{J}$ as
\begin{equation}
    - \mathcal{J} = -\mathscr{D}_\textrm{H$^+$}\frac{\partial\tilde{C}_\textrm{S}}{\partial\tilde{x}}\vert_{\tilde{x}=-1} - \mathscr{D}_\textrm{H$^+$}\tilde{C}_\textrm{S}\frac{\partial\tilde{\phi}}{\partial\tilde{x}}\vert_{\tilde{x}=-1},
    \label{eqn:left_bound_H}
\end{equation}
and
\begin{equation}
    0 = -\mathscr{D}_\textrm{Cl$^-$}\frac{\partial\tilde{C}_\textrm{S}}{\partial\tilde{x}}\vert_{\tilde{x}=-1} + \mathscr{D}_\textrm{Cl$^-$}{}\tilde{C}_\textrm{S}\frac{\partial\tilde{\phi}}{\partial\tilde{x}}\vert_{\tilde{x}=-1}.
    \label{eqn:left_bound_Cl}
\end{equation}
To simplify to a single condition we divide eqns. \ref{eqn:left_bound_H} and \ref{eqn:left_bound_Cl} by $\mathscr{D}_\textrm{H}$ and $\mathscr{D}_\textrm{Cl}$, respectively, and then sum the equations to write
\begin{equation}
    \frac{d\tilde{C_{\textrm{S}}}}{d\tilde{x}}|_{\tilde{x}=-1}=\frac{\mathcal{J}}{2\mathscr{D}_\textrm{H$^+$}}.
    \label{left_bound}
\end{equation}

Similarly, at the anode we write
\begin{equation}
    0 = -\mathscr{D}_\textrm{H$^+$}\frac{\partial\tilde{C}_\textrm{S}}{\partial\tilde{x}}\vert_{\tilde{x}=-1} - \mathscr{D}_\textrm{H$^+$}\tilde{C}_\textrm{S}\frac{\partial\tilde{\phi}}{\partial\tilde{x}}\vert_{\tilde{x}=-1},
    \label{eqn:right_bound_H}
\end{equation}
and
\begin{equation}
    \mathcal{J} = -\mathscr{D}_\textrm{Cl$^-$}\frac{\partial\tilde{C}_\textrm{S}}{\partial\tilde{x}}\vert_{\tilde{x}=-1} + \mathscr{D}_\textrm{Cl$^-$}{}\tilde{C}_\textrm{S}\frac{\partial\tilde{\phi}}{\partial\tilde{x}}\vert_{\tilde{x}=-1},
    \label{eqn:right_bound_Cl}.
\end{equation}
We similarly simplify the equations to write
\begin{equation}
    \frac{d\tilde{C}_\textrm{S}}{d\tilde{x}}|_{\tilde{x}=1}=\frac{-\mathcal{J}}{2\mathscr{D}_\textrm{Cl$^-$}}.
    \label{right_bound}
\end{equation}

At both $\tilde{x} = -1$ and $\tilde{x} = 1$, after $\tau = 0$, $\tilde{C}_{\textrm{S}}$ will decrease from the wall concentration $\tilde{C}_{\textrm{S,wall}}$ to the initial bulk concentration $\tilde{C_{\textrm{S}}}(\tilde{x},\tau =0)= 1$ over a dynamic length scale $\xi(\tau)$. This scenario is indicative of a typical similarity variable solution for diffusion into a semi-infinite domain. In our problem, there are two nuances that we must consider. First, we have diffusion from two boundaries into the bulk of the electrochemical cell. In this way, we can treat our system as a \textit{left} (at the cathode) and \textit{right} (at the anode) seminfinite domain, each with their own similarity variable. This will allow us to solve two separate problems, one for a left salt concentration ($\tilde{C}_{\textrm{S,left}}$) and one for a right salt concentration ($\tilde{C}_{\textrm{S,right}}$). After solving for both concentrations, we then stitch the similarity variable solutions into one solution via $\tilde{C}_{\textrm{S}}(\tilde{x},\
\tau) = \tilde{C}_{\textrm{S,left}}(\tilde{x},\
\tau) + \tilde{C}_{\textrm{S,right}}(\tilde{x},\
\tau) -\tilde{C_{\textrm{S}}}(\tilde{x},\tau =0)$. Secondly, the wall concentration also changes dynamically. This must be incorporated when normalizing, $\tilde{C}_{\textrm{S,p}}$, where $p$ corresponds to either the left or right solution. A similarity variable solution with a dynamically changing reference concentration is discussed in Analysis of Transport Phenomena Example 4.2-2 by William M. Deen \cite{deen_analysis_2012}. This solution will be valid until the middle of the cell has been altered by the concentration profiles from the electrodes, i.e., for $\tau \ll 1$.

\subsection{Similarity Variable Solution}
We begin by assuming a similarity variable of the form
\begin{equation}
    \eta_\textrm{p} = \frac{1 \pm \tilde{x}}{\xi(\tau)}.
    \label{eqn:def_eta}
\end{equation}
For the cathode $\eta_{\textrm{left}}  = \left(1 + \tilde{x}\right)/\xi(\tau)$ and for the anode  $\eta_{\textrm{right}}  = \left(1 - \tilde{x}\right)/\xi(\tau)$. The order of magnitude of the dynamic length scale, i.e. $\mathcal{O}\left(\xi(\tau)\right)$, can be determined via scaling analysis of eqn. \ref{eqn:salt_diffusion} as $\xi(\tau)\sim \sqrt{D_a\tau}$. Lastly, we normalize $\tilde{C}_{\textrm{S,p}}$ by defining $\theta(\eta_\textrm{p})$ as
\begin{equation}
    \theta_p(\eta_p) = \frac{\tilde{C}_{\textrm{S,p}} - \tilde{C_{\textrm{S}}}(\tilde{x},\tau =0)}{\tilde{C_{\textrm{S}}}(\tilde{x},\tau =0) - \tilde{C}_{\textrm{wall,p}}}.
    \label{eqn:def_theta}
\end{equation}
$\tilde{C}_{\textrm{wall,p}}$ can be determined via scaling analysis of eqn. \ref{left_bound} for the cathode or eqn. \ref{right_bound} for the anode as $\tilde{C}_{\textrm{wall,left}}\sim\tilde{C_{\textrm{S}}}(\tilde{x},\tau =0) - \frac{\mathcal{J}}{2D_{\textrm{H}^+}}\sqrt{D_a\tau}$ or $\tilde{C}_{\textrm{wall,right}}\sim\tilde{C_{\textrm{S}}}(\tilde{x},\tau =0) - \frac{\mathcal{J}}{2D_{\textrm{Cl}^-}}\sqrt{D_a\tau}$. Upon suBSAtitution of $\eta_{\textrm{p}}$ and $\theta(\eta_{\textrm{p}})$, taking care to use the chain rule when differentiating, we arrive at the following ordinary differential equation
\begin{equation}
    \frac{d^2\theta_p}{d \eta_p^2}+\frac{\eta_p}{2}\frac{d\theta_p}{d \eta_p}-\frac{\theta_p}{2} = 0
    \label{eqn:sim_ode}.
\end{equation}
Similarly, the boundary conditions can be derived as $d\theta_p/d\eta_p\big{|}_{0} = 1$ and $\theta( \infty) = 0$. We note that this equation applies to both the left and right solutions. The only difference between both solutions are the scalings of $\tilde{C}_{\textrm{wall,p}}$ and $\eta_{\textrm{p}}$. Eqn. \ref{eqn:sim_ode} permits a solution of the form
\begin{equation}
    \theta_p(\eta_p) = \eta_p\textrm{erfc}\left(\frac{\eta_p}{2}\right) + \frac{2}{\sqrt{\pi}}\exp\left(-\frac{\eta_p^2}{4}\right).
    \label{eqn:sim_var_sol}
\end{equation}
Eqns. \ref{eqn:def_eta} and \ref{eqn:def_theta} can then be inserted into eqn. \ref{eqn:sim_var_sol} to determine $\tilde{C}_{\textrm{S,p}}(\tilde{x},\tau)$ and $d\tilde{C}_{\textrm{S,p}}/d\tilde{x}$ upon differentiation. We can then stitch both solutions as described above to find $\tilde{C}_{\textrm{S}}$ and $d\tilde{C}_{\textrm{S}}/d\tilde{x}$. We then use the definition of the total, diffusive, and electromigrative flux for the i$^\textrm{th}$ species as defined in eqn. \ref{eqn:const_flux_nernst} with the definition of $d\tilde{\phi}/d\tilde{x}$ as defined in eqn. \ref{eqn:solve_for_dphi_dx_dim} to solve for the flux profiles.
\subsection{Unsteady State Results}

\begin{figure}[!ht]
\centering
\includegraphics[width=3in]{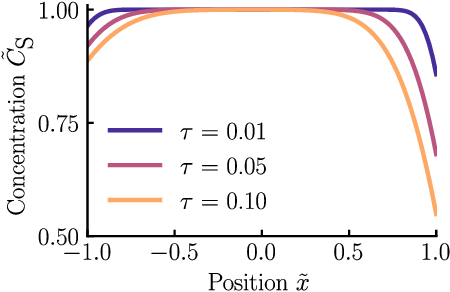}
\caption{\textbf{Transient evolution of salt concentration profiles for hydrolysis of hydrochloric acid electrochemical cell.} The salt concentration profiles as determined by eqn. \ref{eqn:sim_var_sol} for $\tau = 0.01$, $0.05$, and $0.10$ and for $\mathcal{J} = 1$.}
 \label{fig15}
\end{figure}

We begin by evaluating the salt concentration $\tilde{C}_{\textrm{S}}$ throughout the cell for $\tau = 0.01$, $0.05$, and $0.10$, see Fig. \ref{fig15}. As time increases, the depth of penetration of the concentration profile increases, entering the bulk region. Notably, the concentration gradient is larger at the right boundary relative to the left boundary. Upon examination of eqn. \ref{right_bound}, since $\mathscr{D}_\textrm{Cl} = (1/4)\mathscr{D}_\textrm{H}$, the concentration gradient must be larger at the right electrode.

\begin{figure}[!ht]
\centering
\includegraphics[width=6.0in]{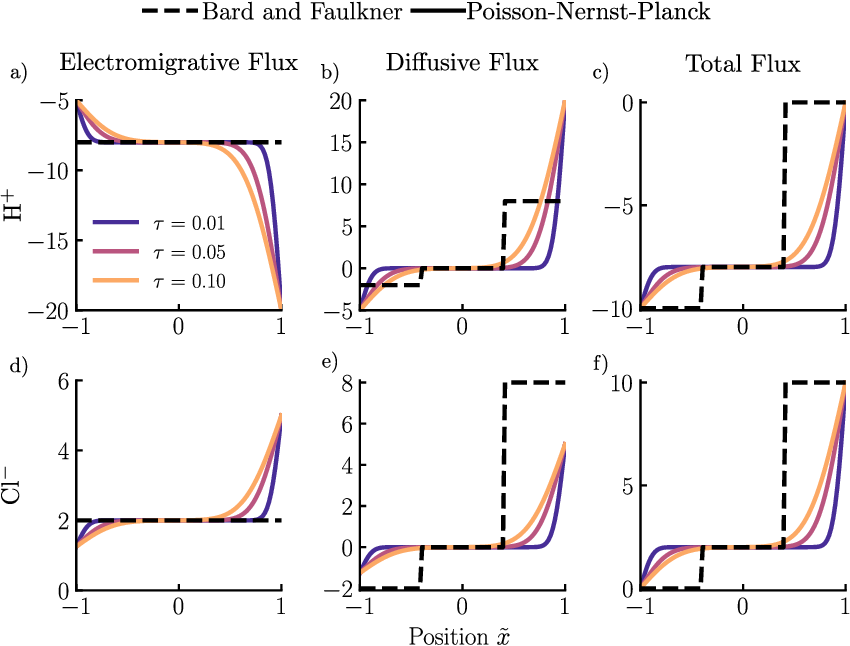}
\caption{\textbf{Transient evolution of flux profiles for H$^+$ and Cl$^-$.} The a,d) electromigrative, b,e) diffusve, and c,f) total flux for H$^+$ and Cl$^-$ as determined by eqn. \ref{eqn:sim_var_sol} for $\tau=\{0.01,0.05,0.1\}$ for $\mathcal{J}=1$ multiplied by $\gamma = 10$ (solid line) and as determined via the balance sheet (dashed line) for $k = 10$ electrons per unit time of current.}
 \label{fig16}
\end{figure}

As the problem is transient, we do see agreement in the bulk between the electromigrative, diffusive, and total fluxes as predicted by the BSA approach, as we saw in section 2.3. This is because the assumption of no concentration gradients is valid. As time increases, electromigrative flux profiles deviate from the constant values in the bulk of -8 and 2 for H$^+$ and Cl$^-$ respectively. Near the electrodes, similar to the examples illustrated in sections 2 and 3, we see that both the the electromigrative and diffusive fluxes are not constant, in stark contrast to the BSA approach. The direction of the overall fluxes can be explained in a similar fashion as the prior examples. While the BSA does qualitatively agree with the PNP equations, and even quantitatively agrees in the bulk for these small times, this agreement will worsen as the electrochemical cell evolves in time. Additionally, as in the previous examples, the BSA poorly predicts the flux contributions near the electrodes and does not provide any information about the concentration profiles. 

\section{Conclusion}
In Electrochemical Methods: Fundamentals and Applications by Alan J. Bard, Larry R.
Faulkner, and Henry S. White, the authors introduce a ``balance sheet" approach to determine the contribution of electromigration and diffusion to the transport of ionic species in an electrochemical cell. While the balance sheet formulation provides a ``simple" means to do such, its application requires the use of a number of assumptions that lack justification. One such assumption is that concentrations are uniform throughout the electrochemical cell except for a small diffusive region near each electrode, isolating diffusive flux to only near the electrodes. The second is that the electromigrative flux is assumed to remain constant across the entire cell, even in the region with a diffusive flux. The consequence of this assumption is that flux conservation for each species is not satisfied. 

In this work, we modeled ion transport in electrochemical cells using PNP equations to determine the contribution of electromigration and diffusion to the flux of ionic species through an electrolyte as well as the species concentration profiles. We first re-worked the 3 examples posed in BFW: a copper redox cell, a copper redox cell with a supporting electrolyte, and a hydrogen evolution cell. We then compared the results from the balance sheet to those predicted by the PNP equations. For the copper redox cell we began by comparing the steady-state PNP equations. Through this example, while we saw agreement with the balance sheet in terms of the direction of the fluxes, quantitative agreement was lacking. The PNP equations predicted concentration profiles that varied throughout the electrochemical cell. This prediction is in disagreement with the assumption of zero bulk concentration gradients as predicted by the balance sheets. Following this, we examined two scenarios where the zero bulk concentration gradients assumption may be valid. First, we examined the solution to the transient PNP equations at short times. Here we saw better agreement in the bulk between the PNP equations and balance sheet approach, however, that agreement was lost near the electrodes and became worse as the system approached steady state. Second, we considered a system where the bulk was continuously stirred, such that the mixing provided a uniform concentration profile through the bulk of the cell except in a small region near the electrodes. We found similar results in this example, qualitatively the balance sheet and PNP equations predict the same direction of fluxes, however, quantitatively they do not agree. Of note for the mixing example, is that due to the presence of mixing, one cannot make claims about the diffusive and electromigrative contributions to flux without conducting a full numerical simulation that incorporates convective species transport. For completeness, after conducting a detailed comparison between the predictions of the balance sheet and PNP equations we evaluate the same copper redox cell but with a supporting electrolyte and the transient evolution of hydrogen and chlorine due to the electrolysis of hydrochloric acid. We scantly saw agreement between the balance sheet approach and predictions of the PNP equations in these examples as well.

 Ultimately, the analysis we present via solution of the PNP equations analysis offers a physically founded route to predict ionic species transport in electrochemical cells. We note that our analysis of the examples was meant to be pedagogical in nature. If used to analyze more realistic scenarios, say for instance in a research setting, key electrochemical phenomena such as reaction kinetics~\cite{bazant2013theory}, double layer dynamics~\cite{jarvey_ion_2022, bazant2004diffuse} and diffusion of electrolytes through porous media~\cite{henrique_parallel_2025,henrique_network_2024,henrique_impact_2022, henrique2022charging, biesheuvel2011diffuse, ratschow2025convection, pedersen2023equivalent}, among others, should be considered. It is our hope that instructors teaching species transport in electrochemical systems will be able to use analyses like those presented in this work, to provide a more physical picture of how ions move in electrochemical systems to their students.

\section*{Acknowledgement}
\noindent RR thanks the National Science Foundation (DGE—2040434) Graduate Research Fellowship for financial support. AG thanks the NSF CAREER award \#2238412 for financial support. The authors thank Adam Holewinski for their useful input on the manuscript. 

\section*{Conflicts of Interest}
\noindent The authors declare no conflicts of interest.

\section*{Data Availability Statement}
\noindent The data that support the findings of this study are available from the corresponding author upon reasonable request.

\section*{Author Contributions}
\noindent \textbf{Grace Origer:} Conceptualization, Writing – original draft, Writing – review and editing \textbf{Ritu R. Raj:} Conceptualization, Writing – original draft, Writing – review and editing \textbf{Nathan Jarvey:}  Writing – review and editing \textbf{P. N. Romero Zavala:} Writing – review and editing \textbf{Wilson A. Smith:} Writing – review and editing \textbf{Ankur Gupta:} Conceptualization, Funding acquisition, Writing – original draft, Writing – review and editing

\bibliographystyle{elsarticle-num} 
\bibliography{references.bib}
\end{document}